\documentclass[fleqn,usenatbib]{mnras}

\usepackage{newtxtext,newtxmath}

\usepackage[T1]{fontenc}
\usepackage{ae,aecompl}

\usepackage{graphicx}	
\usepackage{amsmath}	
\usepackage{amssymb}	
\usepackage{stfloats}  

\title[A $\sim$kpc-scale view of the ISM in `typical' galaxies at $z\sim1.47$]{The kiloparsec-scale gas kinematics in two star-forming galaxies at $z\sim1.47$ seen with ALMA and VLT-SINFONI}

\author[Molina, J. et al.]{
J. Molina,$^{1}$\thanks{E-mail: jumolina@das.uchile.cl}
Edo~Ibar,$^{2}$
I.~Smail,$^{3}$
A.~M.~Swinbank,$^{3}$
E. Villard,$^{4}$
A.~Escala,$^{1}$
D.~Sobral,$^{5}$
\newauthor{T.~M.~Hughes$^{2,6,7,8}$}
\\
$^{1}$Departamento de Astronom\'ia, Universidad de Chile, Casilla 36-D, Santiago, Chile\\
$^{2}$Instituto de F\'isica y Astronom\'ia, Universidad de Valpara\'iso, Avda. Gran Breta\~na 1111, Valpara\'iso, Chile\\
$^{3}$Centre for Extragalactic Astronomy, Department of Physics, Durham University, South Road, Durham DH1 3LE, UK\\
$^{4}$Joint ALMA Observatory/ESO, Avenida Alonso de C\'ordova 3107, Vitacura, Santiago, Chile\\
$^{5}$Department of Physics, Lancaster University, Lancaster, LA1 4YB, UK \\
$^{6}$CAS Key Laboratory for Research in Galaxies and Cosmology, Department of Astronomy, University of Science and Technology of China, Hefei 230026, China\\
$^{7}$School of Astronomy and Space Science, University of Science and Technology of China, Hefei 230026, China\\
$^{8}$Chinese Academy of Sciences South America Center for Astronomy, China-Chile Joint Center for Astronomy, Camino El Observatorio \#1515, Las Condes, Santiago, Chile\\
}

\date{Accepted XXX. Received YYY; in original form ZZZ}

\pubyear{2019}

\begin{document}
\label{firstpage}
\pagerange{\pageref{firstpage}--\pageref{lastpage}}
\maketitle

\begin{abstract}
We present Atacama Large Millimeter/submillimeter Array (ALMA) CO($J=2-1$) observations of two main-sequence
star-forming galaxies at $z\sim1.47$ taken from the High-Z Emission Line Survey (HiZELS). These two systems have been previously 
reported to be molecular gas rich $f_{\rm H_2} \equiv M_{\rm H_2}/(M_{\rm H_2} + M_\star) \sim 0.8$. We carried out a follow-up study 
to resolve, at $\sim$kpc-scales, the CO emission. These new observations are combined with our earlier ALMA observations (sensitive 
to diffuse CO emission) and compared with our previous H$\alpha$-based study at matched spatial resolution. One target 
is marginally resolved in CO(2-1), showing complex dynamics with respect to the ionised gas traced by H$\alpha$. While the other source 
is spatially resolved, enabling a detailed exploration of its internal dynamical properties. In this system, both gaseous 
phases show similar spatial extension, rotation velocities and velocity dispersions ($V_{\rm rot} \sim \sigma_v \sim 100$\,km\,s$^{-1}$) 
suggesting a rotational velocity to velocity dispersion ratio consistent with unity. By comparing the ionized and molecular gas tracers through 
the use of a two-dimensional kinematic model, we estimate a median depletion time $\tau_{\rm dep}=2.3 \pm 1.2$\,Gyr for the galaxy as a 
whole. This value is in agreement with the average $\tau_{\rm dep}$ value observed in local star-forming galaxies at similar 
spatial scales. Using a thick-disk dynamical modelling, we derive a dynamical mass $M_{\rm dyn} = (1.59\pm0.19) \times 10^{11}$\,$M_\odot$ 
within $\approx 6$\,kpc. This suggests a dark matter fraction ($f_{\rm DM} \equiv M_{\rm DM}/M_{\rm dyn}$) of $0.59\pm0.10$, in agreement 
with the average $f_{\rm DM}$ value derived from stacked rotation curve analysis of galaxies at similar redshift range.
\end{abstract}




\section{Introduction}

Understanding how galaxies form and evolve over cosmic time is a major
goal in modern astrophysics. Surveys have shown that there is a decline
in the overall cosmic star-formation rate density since $z \sim 2$ (e.g. 
\citealt{Madau1996,Sobral2013,Khostovan2015}) which coincides with the 
decrease of the average fraction of molecular gas mass in galaxies (e.g. 
\citealt{Tacconi2010,Geach2012,Carilli2013}). This behaviour is thought to 
match the cosmic evolution of the mass in stars, and the molecular gas 
content ($M_{\rm H_2}$) of the Universe, hence it provides a logical 
interpretation for the interplay between, perhaps, the main actors 
controlling the growth of galaxies (e.g. \citealt{Madau2014}). 

At the peak epoch of the cosmic star formation activity ($z\sim2-3$), spatially-resolved observations 
of galaxies have mostly come from large \textit{Hubble Space Telescope} (\textit{HST}) 
and Integral Field Unit (IFU) surveys (e.g. \citealt{Koekemoer2011,Brammer2012,Law2012}). The 
latter trace the ionized gas content in seeing limited conditions ($\sim 0\farcs6$ in $K-$band, e.g. 
\citealt{Sobral2013b,Wisnioski2015,Stott2016,Turner2017,Johnson2018}). Although adaptive optics 
(AO)-aided IFU observations have delivered $\sim 0\farcs15$ ($\sim$kpc-scale) spatial resolution 
data on smaller galaxy samples (e.g. \citealt{Forster2009,Swinbank2012a,Molina2017,Forster2018,Gillman2019}).
Deep observations have focused mainly in sampling the `main-sequence' of star-forming 
galaxies, i.e., those galaxies that are part of the bulk of the galaxy population in terms of stellar mass 
($M_\star$) and star formation rate (SFR; e.g. \citealt{Noeske2007,Whitaker2012}).

High redshift ($z \sim 1-3$) IFU surveys targeting the H$\alpha$ emission have revealed that most of 
the main-sequence star-forming galaxies (hereafter, `typical' star-forming galaxies), present: 
(1) highly turbulent galactic disks with high surface brightness, indicating that the interstellar medium 
(ISM) is highly pressurized with $P_{\rm tot} \sim 10^{3-4}$ times higher than the typical ISM pressure
in the Milky Way \citep{Swinbank2015,Molina2017}; (2) the star-formation activity is partly triggered by 
gravitational fragmentation of dynamically unstable gas potentially leading to the formation of massive 
clumps which could be up to $\sim 1000 \times$ more massive ($\sim 10^9$\,$M_\odot$) than star-forming 
complexes seen in local galaxies (e.g. \citealt{Genzel2011,Swinbank2012b}).

\begin{table*}
\centering
TABLE 1: ALMA OBSERVATIONAL SETUP\\
\caption{\label{tab:table1}
ALMA Cycle-5 observations. These data have been concatenated with the data shown in Hughes et al. (in prep.).
 `PWV' is the average precipitable water vapour estimate for the observations}
\begin{tabular}{lcccccccc}
\hline
Source & Project ID & Observation & Flux & Bandpass & Phase & PWV & Number of & Time on \\
 & & Date & Calibrator & Calibrator & Calibrator & (mm) & antennas & Target (min)\\
\hline
SHiZELS-8 & 2017.1.01674.S & 14 November 2017 & J0238+1636 & J0238+1636 & J0217-0820 & 3.17 & 43 & 47.05 \\
 & & 15 November 2017 & J0238+1636 & J0238+1636 & J0217-0820 & 2.05 & 43 & 45.78 \\
&  & 16 November 2017 & J0006-0623 & J0006-0623 & J0217-0820 & 1.44 & 43 & 45.82 \\
\hline
COS-30 / SHiZELS-19 & 2017.1.01674.S & 14 November 2017 & J1058+0133 & J1058+0133 & J0948+0022 & 3.56 & 43 & 43.67 \\
 & & 14 November 2017 & J1058+0133 & J1058+0133 & J0948+0022 & 3.92 & 43 & 44.12 \\
 & & 16 November 2017 & J1058+0133 & J1058+0133 & J0948+0022 & 0.89 & 43 & 44.23 \\
 & & 18 November 2017 & J1058+0133 & J1058+0133 & J0948+0022 & 0.60 & 43 & 44.50 \\
 & & 20 November 2017 & J1058+0133 & J1058+0133 & J0948+0022 & 0.48 & 50 & 44.02 \\
\hline
\end{tabular}
\end{table*}
 
Although the physical conditions that produce these extreme ISM properties 
remain poorly understood, one possible explanation may be related to the 
high molecular gas densities that may arise from the high molecular gas 
fractions ($f_{\rm H_2}$; e.g. \citealt{Escala2008}). In the local Universe galaxies 
have typical $f_{\rm H_2}$ values of $\sim 0.1$, while on 
the other hand galaxies at $z\sim1-3$ have reported molecular gas fractions up to 
$\sim0.8$ (e.g. \citealt{Tacconi2010, Daddi2010}; Hughes et al. in prep.). 
The molecular gas content seems to dominate the baryonic mass budget in 
the central parts of these high redshift `typical' star-forming systems, but we have 
little or almost no information about their spatial distribution and kinematics.

Traditionally the workhorse tracer to estimate the 
molecular gas content are the low$-J$ rotational transitions of the 
carbon monoxide ($^{12}$C$^{16}$O) molecule (e.g. $J = 1 - 0$ or $J = 2 - 1$; 
hereafter CO(1-0) and CO(2-1), respectively; \citealt{SV2005, Bolatto2013}). 
Through the assumption of a CO-to-H$_2$ conversion factor ($\alpha_{\rm CO}$), 
the molecular gas to CO(1-0) luminosity ($L'_{\rm CO(1-0)}$) relation can be expressed as 
$M_{\rm H_2}=\alpha_{\rm CO}L'_{\rm CO(1-0)}$ (e.g. \citealt{Bolatto2013}). 
In the Milky Way and other `normal' star-forming local galaxies, the CO emission 
mainly arises from individual virialized Giant Molecular Clouds (GMCs). On the 
other hand, the CO emission coming from more extreme star-forming and 
dynamically disrupted systems, such as Ultra Luminous Infra-red Galaxies (ULIRGs;
\citealt{Downes1998}) is likely to be contained in much denser rotating disks or rings \citep{SV2005}.

Spatially resolved morpho-kinematic studies of the molecular gas content in galaxies 
are critical to understand the physical processes that control the star formation activity.
Nevertheless, observations of high redshift galaxies with direct spatially resolved 
molecular gas detections have remained a challenge. Beyond the local Universe, resolved 
CO detections are limited to the most massive/luminous yet rare galaxies or highly 
magnified gravitationally lensed sources (e.g. \citealt{Saintonge2013,Swinbank2015,Chen2017,
Calistro2018, Motta2018}). With ALMA, we are now able to study the physical conditions 
of the cold molecular gas in `typical' star-forming galaxies at $z > 1$ and test if the actual 
cosmological models successfully explain the characteristics of the high redshift ISM.

In this paper, we use high angular resolution ALMA observations to 
characterize the CO(2-1) emission and kinematics of two `typical' galaxies (following 
the so-called `main-sequence') at $z \sim 1.47$ drawn from the SHiZELS survey 
(\citealt{Swinbank2012a,Molina2017, Gillman2019}). Combining ALMA with the 
available AO-aided H$\alpha$ data observed by the Spectrograph for 
INtegral Field Observations in the Near Infrared (SINFONI) on the Very Large 
Telescope (VLT), we study how the spatially-resolved properties of the ionized 
and cold molecular gas are related on $\sim$kpc-scales. Throughout the paper, we 
adopt a $\Lambda$CDM cosmology with $\Omega_{\Lambda}$=0.73, 
$\Omega\rm_m$=0.27, and H$_0$=70 km\,s$^{-1}$\,Mpc$^{-1}$, implying a spatial 
resolution of $\approx 0\farcs15$ that corresponds to a physical scale of $\sim 1$\,kpc. 
We assume a \citet{Chabrier2003} Initial Mass Function (IMF) and a Solar Oxygen 
abundance of 8.69$\pm$0.05 in the 12+$\log_{10}$(O/H) metallicity scale \citep{Asplund2009}.

\section{Observations \& Data Reduction}
\subsection{The SHiZELS Survey} 
\label{sec:shizels}

In this work, we take advantage of galaxies with previous H$\alpha$ VLT-SINFONI IFU 
AO-aided imaging taken from the SHiZELS survey (\citealt{Swinbank2012a,Molina2017,Gillman2019}). 
This is based on a sub-sample of sources taken from the HiZELS near-infrared 
narrow-band imaging project \citep{Sobral2012,Sobral2013,Sobral2015} and is 
one of the largest IFU-AO survey observing the H$\alpha$ emission in `typical' 
star-forming galaxies on $\sim$kpc-scales in three narrow redshift slices $z = 0.86$, 1.47 \& 2.23
($M_\star \sim 3 - 30 \times 10^{10}$\,$M_\odot$; SFR$\sim 2-30$\,$M_\odot$\,yr$^{-1}$). 
All galaxies have a deep multi-wavelength coverage as they line within the UDS, 
COSMOS and SA22 fields.

From SHiZELS, we select two galaxies, COS-30 and SHiZELS-8,
which have been previously detected in CO($J=2-1$) with ALMA
at $\sim1\farcs6-2\farcs5$ resolution (Hughes et al. in prep.).

\begin{figure*}
\includegraphics[width=0.45\columnwidth]{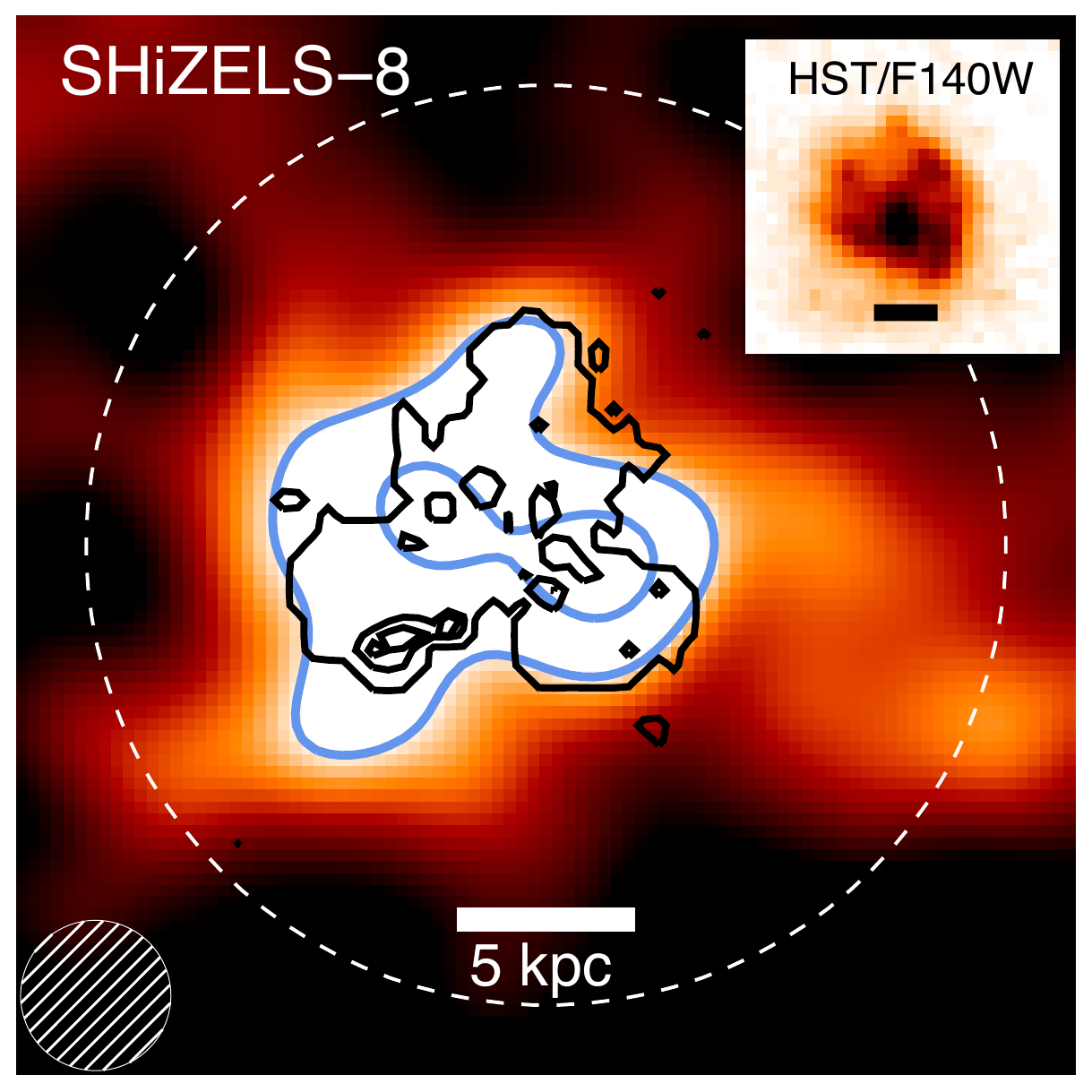}
\includegraphics[width=0.56\columnwidth]{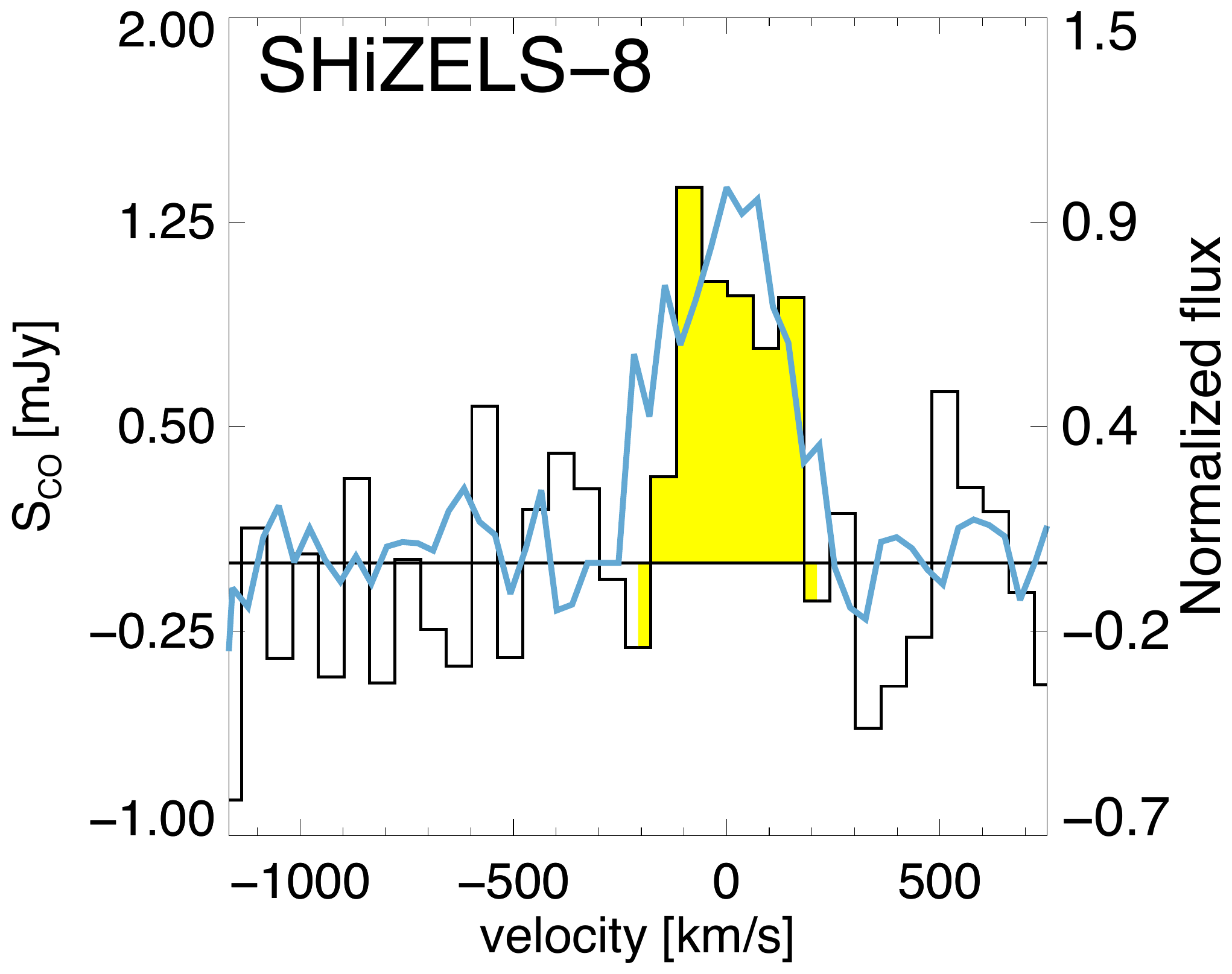}
\hspace{3mm}
\includegraphics[width=0.45\columnwidth]{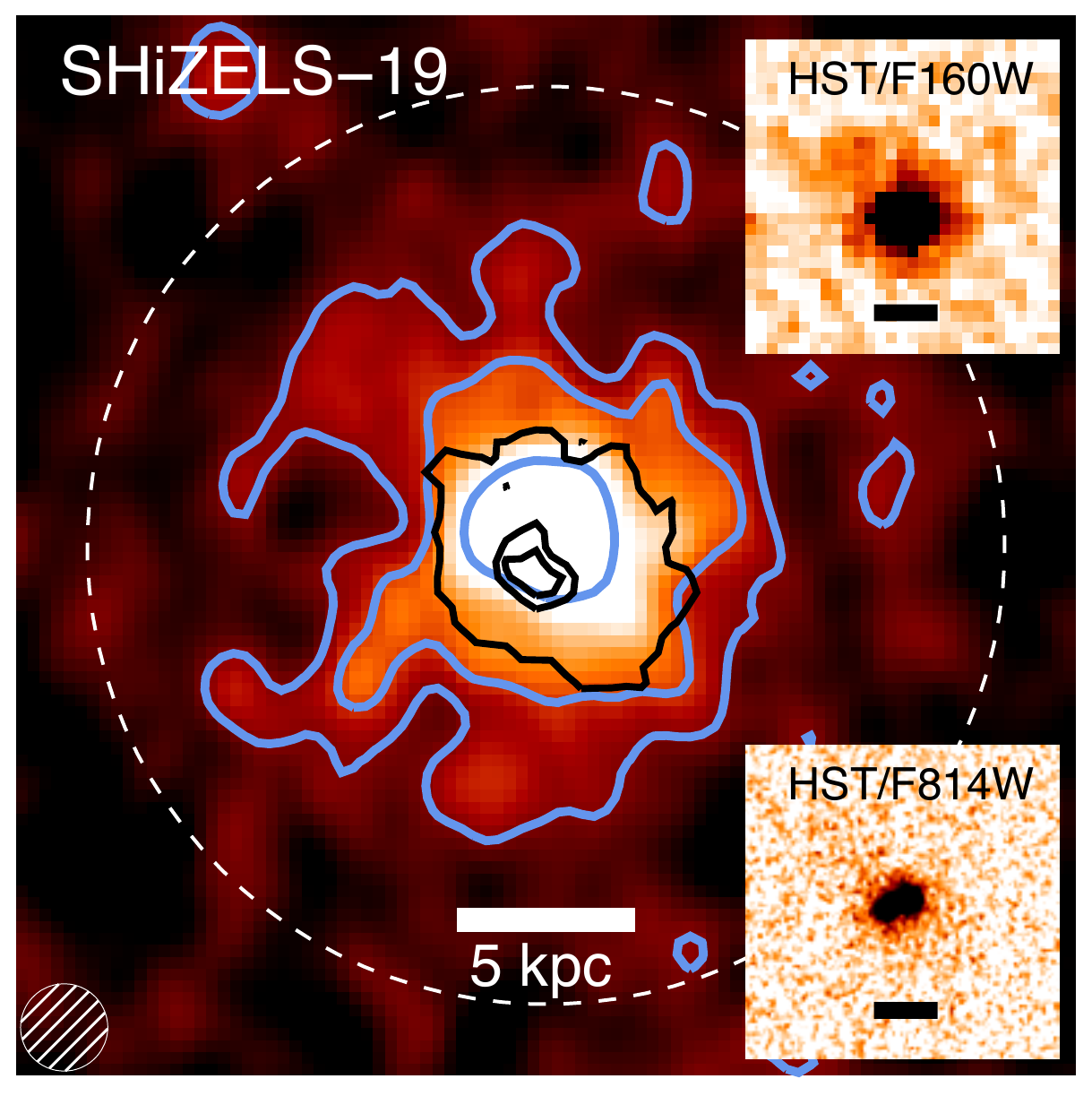}
\includegraphics[width=0.53\columnwidth]{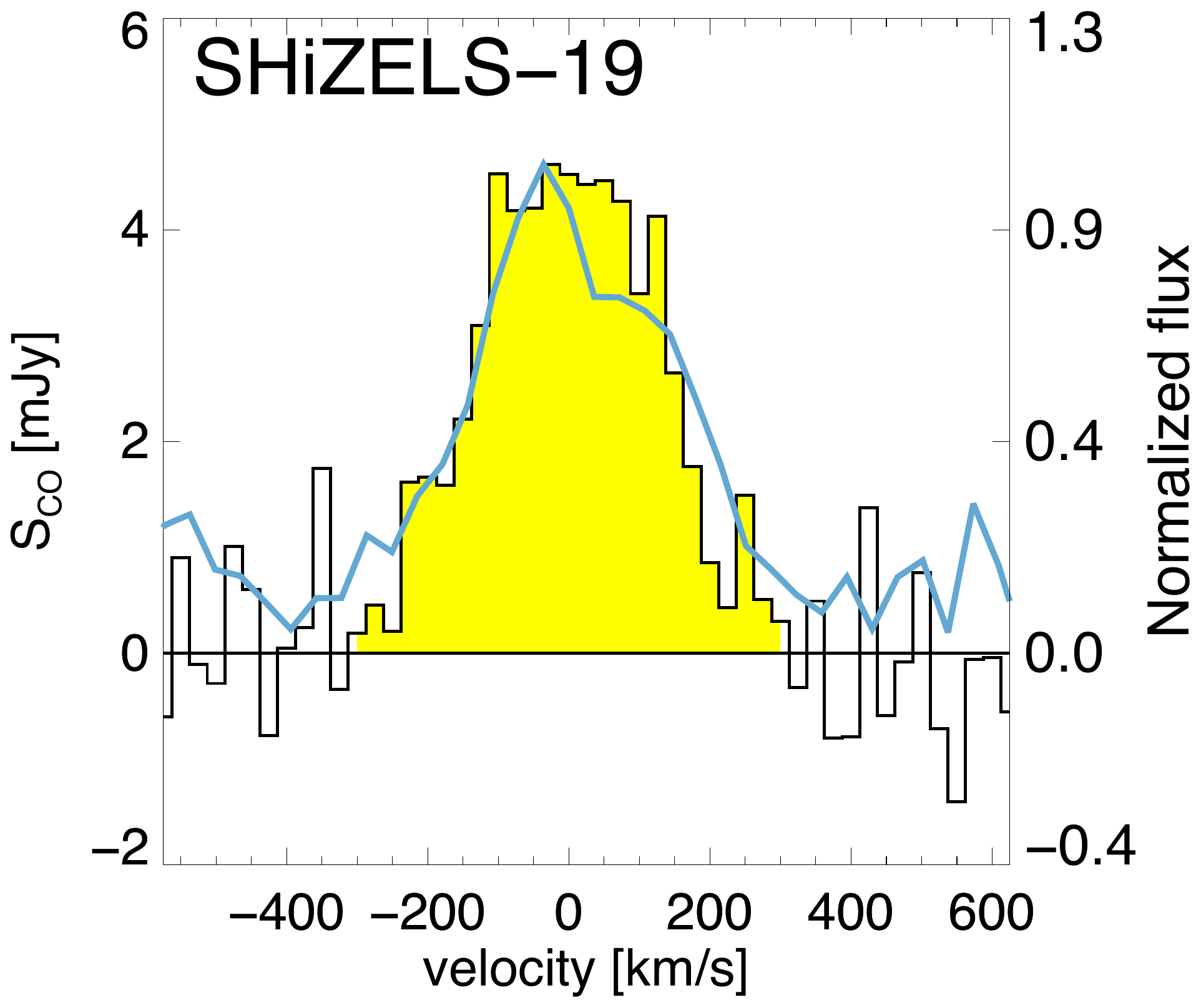}
\caption{ \label{fig:fig1}
\textit{1st and 3rd columns}: Spectrally integrated 2000\,k$\lambda$ data-cubes encompassing the 
CO(2-1) emission line for each galaxy in a $3\farcs56$ ($\approx 30$\,kpc) squared sky 
region. The synthesized beam size ($\theta_{\rm BMAJ} = 0\farcs50$, 0$\farcs$29 for 
SHiZELS-8 and SHiZELS-19, respectively) is showed in the bottom-left corner in each map. 
The blue contours represent the 3-, 5- and 10-$\sigma$ levels
the image noise. The black contours show the H$\alpha$ emission detected in the
SINFONI-AO observations. For SHiZELS-19 we align both intensity maps by using their 
best-kinematic centres (see \S~\ref{sec:galaxy_dyn}). In contrast, as we lack the detailed 
kinematic information for SHiZELS-8, we just align its intensity maps by eye in order to 
improve visualization. The dashed line represents the sky-aperture defined as 
2$\times$\,FWHM of the best-fitted two-dimensional Gaussian in each map. If 
available, we also show the \textit{HST} broad-band images over the same sky region in the right 
side of the map. In each \textit{HST} cut-out, the black bar represents the 5\,kpc scale.
\textit{2nd and 4th columns:} Spatially-collapsed spectra extracted within the 
sky-aperture for each galaxy showing the CO(2-1) emission line. The yellow colour indicates 
the $2\times$\,FWHM region for the CO emission line. The blue line shows the H$\alpha$ emission 
line flux density normalized to the CO emission line peak and extracted from the 
SINFONI-AO IFU observations using the same sky aperture \citep{Swinbank2012a,Molina2017,Gillman2019}.
We find good agreement between the CO and H$\alpha$ line widths.
}
\end{figure*}

The global stellar masses and SFRs are taken from \citet{Gillman2019}.
Briefly, the stellar masses were computed by using the Bayesian SED fitting 
code, {\sc MAGPHYS} \citep{daCunha2008} to the rest-frame UV, optical 
and near-infrared data available ($FUV,NUV,U,B,g,V,R,i,I,z,Y,J,H,K,3.6,4.5,5.8$ 
and 8.0\,$\micron$ collated in \citealt{Sobral2014}, and references therein), 
assuming a \citet{Chabrier2003} IMF and a \citet{Calzetti2000} extinction law.

The SFRs are calculated from the $M_\star$-based extinction-corrected 
H$\alpha$ emission line fluxes \citep{Garn2010, Sobral2012, Ibar2013} and adopting 
the \citet{Kennicutt1998a} calibration SFR$_{\rm H\alpha}$\,($M_\odot$\,yr$^{-1}$)= 
$4.6 \times 10^{-42}$\, L$_{\rm H\alpha}$\,(erg\,s$^{-1}$) with a \citet{Chabrier2003} 
IMF. The total H$\alpha$ emission line fluxes are taken from the HiZELS narrow-band 
survey and are corrected for [N{\sc ii}] flux contamination by considering the
[N{\sc ii}]/H$\alpha$ ratio measured from the SINFONI observations. We note that the 
extinction corrected SFR values presented in this work are consistent with the values
reported in \citet{Gillman2019}, i.e, with the intrinsic SFR values estimated by {\sc MAGPHYS}. 
We note that the COS-30 galaxy is referred to as `SHiZELS-19' in \citet{Gillman2019}. 
Hereafter we use this name to refer to this galaxy.

We adopt the \citet{Whitaker2012}'s definition of the `main-sequence' of star-forming galaxies, and 
by using the redshift, $M_\star$ and the specific star formation rate (sSFR$\equiv$SFR/$M_\star$)
estimates for each source, we calculate the `distance' to the main-sequence 
($\Delta$MS$\equiv$sSFR/sSFR$_{\rm MS}$($z$,$M_\star$)). We present the $\log_{10}$($\Delta$MS) 
values in Table~\ref{tab:table2}. These values are lower than the 0.6\,dex upper limit usually adopted to 
define the main-sequence \citep{Genzel2015}.

\subsection{ALMA observations \& data reduction}
\label{sec:alma_datared}
We made use of Cycle-5 ALMA Band-3 observations (2017.1.01674.S;\,P.I.\,Molina J.; 
see Table~\ref{tab:table1}) to detect and resolve the redshifted CO(2--1) emission line 
($\nu_{\rm rest} = 230.538$\,GHz) for SHiZELS-8 \citep{Swinbank2012a} and 
SHiZELS-19 (presented as COS-30 in \citealt{Molina2017}). Those observations were carried 
during November 2017, reaching a root-mean-squared (r.m.s.) noise of 120--150\,$\mu$Jy\,beam$^{-1}$
 at 0$\farcs$15 angular resolution using a channel width of 60\,km\,s$^{-1}$.

The Cycle-5 observations were taken in an extended configuration (synthesized beam 
FWHM of $\approx 0\farcs15$), thus being more sensitive to more compact emission. 
We combine them with previous $2''$ resolution Cycle-1 and -3 ALMA data (see Hughes et al. 
in prep. for more details) to obtain sensitive and high-fidelity imaging of the CO(2-1) emission.

Data were reduced using {\sc Common Astronomy Software Applications}\footnote{http://casa.nrao.edu/index.shtml} ({\sc casa}) 
considering a standard ALMA pipeline up to calibrated $u$-$v$ products. 
We used the task {\sc tclean} to deconvolve the data to produce datacubes for both 
galaxies. In each datacube we clean the regions where emission is identified down to 3-$\sigma$ 
using the {\sc tclean} {\sc casa} task, allowing multi-scale cleaning ({\sc multiscale=[0.5,5,15,45]}, 
where image pixel size is fixed at 0$\farcs$04). The high-resolution datacubes are produced by 
using Briggs weighting with robust parameter at 0.5, obtaining synthesized beam FWHMs of 
$\approx 0\farcs15$ ($\sim$\,kpc-scale at $z \sim 1.47$). We also take advantage of the Cycle-1 
and -3 data by producing datacubes with different spatial scales by tapering at 2000\,k$\lambda$ 
and reducing the spatial resolution using a circular restoring synthesized beam ($0\farcs29 \approx 
2.5$\,kpc at $z \sim 1.47$). These combined tapered datacubes are produced with the aim of recovering 
as much as the low surface brightness CO(2-1) emission as possible from the outskirts of each galaxy (Fig.~\ref{fig:fig1}).

In the case of SHiZELS-8 we are unable to detect the CO emission in the high-resolution datacube or the $\approx 2.5$\,kpc
resolution map. Therefore, for this galaxy, we reduce the spectral and spatial resolutions in order to 
boost the CO emission signal-to-noise (S/N). The spectral channel width is set to 60\,km\,s$^{-1}$ and the spatial 
resolution is degraded to $0\farcs50$ by performing an additional smoothing step. 

In the case for SHiZELS-19, we are able to easily detect the source in the high-resolution datacube. Thus, for this galaxy, 
we set the spectral channel width to 25\,km\,s$^{-1}$ aiming to minimize spectral resolution effects in the derivation of the 
kinematic parameters.

We show the spatially integrated spectrum for each galaxy in Fig.~\ref{fig:fig1}. Those were extracted by considering 
a sky-aperture defined in diameter as 2$\times$\,FWHM of the best-fitted two-dimensional Gaussian in each map
($\sim 1\farcs3-1\farcs2$ for SHiZELS-8 and SHiZELS-19).

In summary, for SHiZELS-19 we combine Cycle-1, -3 and -5 data to generate a high resolution ($\approx 0\farcs15 \sim$kpc-scale) 
and a `low resolution'  ($\approx 0\farcs29 \sim 2.5$\,kpc) datacubes, while for SHiZELS-8 we use
a $\approx 0\farcs5$ resolution map ($\sim 4.3$\,kpc), optimizing the flux
sensitivity to the compact and diffuse emission in each source, respectively.

\begin{table}
	\centering
	\setlength\tabcolsep{4pt}
	TABLE 2: SPATIALLY-INTEGRATED GALAXY PROPERTIES\\
    	\caption{\label{tab:table2}                                                   
	The integrated H$\alpha$ flux densities (f$_{\rm H\alpha}$) are taken from narrow-band 
	photometry and corrected for [N{\sc ii}] contamination. The SFR$_{\rm H\alpha}$ values are corrected 
	for H$\alpha$ extinction ($A_{\rm H\alpha}$) following the $M_\star-A_{\rm H\alpha}$ parametrization 
    presented by \citet{Garn2010}. $\Delta$MS is the offset of each galaxy with respect to the 
    `main-sequence' of star-forming galaxies. $\alpha_{\rm CO,A+17}$ and $\alpha_{\rm CO,N+12}$ are 
    the CO-to-H$_2$ conversion values calculated by following the \citet{Accurso2017} and 
    \citet{Narayanan2012} parametrizations. The $M_{\rm H_2}$ and $f_{\rm H_2}$ quantities are computed 
    by using $\alpha_{\rm CO,N+12}$ (see \S~\ref{sec:dyn_mass}).}
	\begin{tabular}{lcc} 
		\hline
		\hline
		ID & SHiZELS-8 & SHiZELS-19 \\
        \hline
        RA (J2000) & 02:18:21.0 & 09:59:11.5 \\
        Dec (J2000) & $-$05:19:07.8 & +02:23:24.3\\
        $z_{\rm spec}$ & 1.4608 & 1.4861 \\
        f$_{\rm{H\alpha}}/10^{-17}$ (erg\,s$^{-1}$\,cm$^{-2}$) & 10.9$\pm$1 & 7.6$\pm$1\\
        $A_{\rm H \alpha}$ & 1.1$\pm$0.2 & 1.1$\pm$0.2\\
        $[$N{\sc ii}$]$/H$\alpha$ & <0.1& 0.43$\pm$0.03 \\
        SFR$_{\rm H\alpha}$ ($M_{\odot}$\,yr$^{-1}$) & 16$\pm$2 & 13$\pm$2 \\
        $\log_{10} M_\star$ ($M_{\odot}$) & 10.3$\pm$0.2 & 10.3$\pm$0.2  \\
        $\log_{10} \Delta$MS (dex) & 0.53 & 0.41\\
        $S_{\rm CO(2-1)} \Delta v$ (Jy\,km\,s$^{-1}$) & 0.38$\pm$0.08 & 0.64$\pm$0.03 \\
        $\log_{10} L'_{\rm{CO(2-1)}}$ (K\,km\,s$^{-1}$\,pc$^2$) & 10.04$\pm$0.04 & 10.27$\pm$0.04 \\
        $\alpha_{\rm CO,A+17}$ ($M_{\odot}$(K\,km\,s$^{-1}$\,pc$^2$)$^{-1}$) & 21$\pm$8 & 3.9$\pm$1.5 \\
        $\alpha_{\rm CO,N+12}$ ($M_{\odot}$(K\,km\,s$^{-1}$\,pc$^2$)$^{-1}$) & 5.0$\pm$1.0 &  1.5$\pm$0.2 \\
        $\log_{10} M_{\rm H_2}$ ($M_{\odot}$) & 10.81$\pm$0.10 & 10.51$\pm$0.07 \\
        $f_{\rm H_2}$ & 0.76$\pm$0.24 & 0.62$\pm$0.16 \\
		\hline                                                                                        
	\end{tabular}
\end{table}

\section{ANALYSIS, RESULTS \& DISCUSSION}

\subsection{CO emission \& CO-to-H$_2$ conversion factor}
\label{sec:molgas_cont}
The global CO(2-1) velocity-integrated flux densities ($S_{\rm CO(2-1)} \Delta v$) are taken from 
Hughes et al. (in prep.) and presented in Table~\ref{tab:table2}. Those are estimated 
by fitting a 2D Gaussian profile to the spectrally-integrated datacube (moment 0).
The CO(2-1) luminosities ($L'_{\rm CO(2-1)}$) are calculated by following \citet{SV2005};

\begin{equation}
L'_{\rm CO(2-1)} = 3.25 \times 10^7\,S_{\rm CO(2-1)} \Delta v\, \nu_{\rm obs}^{-2}\, D^2_{\rm L}\, (1+z)^{-3}\, {\rm [K\,km\,s^{-1}\,pc^2]}, 
\end{equation}

\noindent where $S_{\rm CO(2-1)} \Delta v$ is in Jy\,km\,s$^{-1}$, $\nu_{\rm obs}$ 
is the observed frequency of the emission line in GHz, D$_{\rm L}$ is the luminosity 
distance in Mpc, and $z$ is the redshift. We then estimate the CO(1-0) luminosity for 
each galaxy by assuming a $L'_{\rm CO(2-1)}/L'_{\rm CO(1-0)} = 0.85$ ratio 
(e.g. \citealt{Danielson2011}). 

To derive molecular gas masses we need to assume a CO-to-H$_2$ conversion factor. By considering 
a dynamical model we constrain the $\alpha_{\rm CO}$ value in our galaxies (see \S~\ref{sec:dyn_mass}). However, 
we also use different prescriptions in the literature to calculate tentative CO-to-H$_2$ conversion factor values.
Unfortunately, as we lack of dust masses for SHiZELS-8 and SHiZELS-19 (see Cheng et al. in prep), 
we are unable to use a dust-to-gas ratio motivated $\alpha_{\rm CO}$ value (e.g. \citealt{Leroy2013}).
Thus, from the literature we use the \citet{Accurso2017} and \citet{Narayanan2012} 
$\alpha_{\rm CO}$ prescriptions as we have direct estimates of the input observables and
these parametrizations do not require a minimum observational spatial resolution (e.g. \citealt{Feldmann2012}). 

Briefly, \citet{Accurso2017}'s prescription considers the effect of the ISM metallicity and the strength 
of the UV radiation field in the estimation of the CO-to-H$_2$ conversion factor. We note that in this 
parametrization, the strength of the UV field is traced by the offset of the galaxy with respect to the 
`main-sequence' of star-forming galaxies ($\Delta$MS; see \citealt{Accurso2017}, for more details). 
However, this prescription does not consider deviations of the $\alpha_{\rm CO}$ value due to high 
gas surface density ($\Sigma_{\rm gas}$) values (e.g. \citealt{Bolatto2013}). In contrast, the 
\citet{Narayanan2012}'s prescription takes into account the effect of the ISM metallicity and gas 
surface density in the estimation of the $\alpha_{\rm CO}$ value. This is, however, a numerical 
prediction for $\Sigma_{\rm gas}$ and its effect is parametrized via the luminosity-weighted CO 
surface brightness quantity ($\Sigma_{\rm CO}$; see \citealt{Narayanan2012}, for more details).

In order to apply these two $\alpha_{\rm CO}$ parametrizations, we use the $\Delta$MS values
calculated by assuming the \citet{Whitaker2012}'s definition of the `main-sequence' of star-forming galaxies 
and presented in Table~\ref{tab:table2}. The metallicities are estimated 
from the [N{\sc ii}]/H$\alpha$ ratio and assuming the \citet{Pettini2004} metallicity prescription. The inclination 
corrected $\Sigma_{\rm CO}$ values are calculated from the ALMA observations. Based on these assumptions, 
we list the global $\alpha_{\rm CO}$ values for each galaxy in Table~\ref{tab:table2}. 

We find little agreement between the two parametrizations. By considering the \citet{Accurso2017}'s prescription, 
we find higher CO-to-H$_2$ conversion values than the obtained from the \citet{Narayanan2012}'s parametrization 
(Table~\ref{tab:table2}). This is expected as \citet{Accurso2017}'s prescription does not consider the effect of 
$\Sigma_{\rm gas}$ in their estimation of the $\alpha_{\rm CO}$, and it has a steeper dependence on metallicity.
In the case of the \citet{Narayanan2012}'s parametrization, the low $\alpha_{\rm CO}$ value obtained for SHiZELS-19 is 
mainly dominated by its high galactic $\Sigma_{\rm CO}$, which is reflected by its high $\Sigma_{\rm H_2}$ 
value (Table~\ref{tab:table4}). On the other hand, SHiZELS-8 has an $\alpha_{\rm CO}$ value closer to 
that found in Galactic GMCs (Table~\ref{tab:table2}). This is produced by its low (sub-solar) metal content 
(12+$\log_{10}$(O/H)\,<\,8.12). Although variations of the CO-to-H$_2$ conversion factor within galactic disks have 
been reported (e.g. \citealt{Sandstrom2013}), we note that a global $\alpha_{\rm CO}$ value seems to be a good 
approximation for the SHiZELS-19 galaxy (Appendix~\ref{appendix:A}). 

\subsection{The SHiZELS-8 galaxy}
\label{sec:SHiZELS-8}

The SHiZELS-8 H$\alpha$ observation \citep{Swinbank2012a} suggests that this galaxy is
consistent with being a turbulent rotating disk hosting three kpc-sized clumps \citep{Swinbank2012b}.
Unfortunately the SHiZELS-8 CO(2-1) observations have too low S/N to allow a detailed dynamical 
characterization. This galaxy has $\sim$50\% lower velocity-integrated CO(2-1) flux density than 
SHiZELS-19, but its emission seems more extended, i.e, it has a lower CO surface brightness. 
On the other hand, our estimated low metallicity for SHiZELS-8 (12+$\log_{10}$(O/H)\,<\,8.12) suggests a 
lack of dust content which could indicate an efficiently CO molecule photo-dissociation by the far-ultraviolet 
(far-UV) photons and a higher CO-to-H$_2$ conversion factor \citep{Bolatto2013}. This implies that 
SHiZELS-8 could have a larger molecular gas content than SHiZELS-19, albeit similar SFR 
and $M_\star$ (see Table~\ref{tab:table2}). 

From the high-resolution ($0\farcs15 \sim$kpc-scale) datacube we obtain a velocity-integrated 
peak flux density r.m.s of 3.4\,mJy\,km\,s$^{-1}$\,beam$^{-1}$, corresponding a $\Sigma_{\rm H_2}$ upper 
limit of $\sim 1.6\times10^3$\,$M_\odot$\,pc$^{-2}$\,beam$^{-1}$ based on the \citet{Narayanan2012}'s 
CO-to-H$_2$ conversion factor (see \S~\ref{sec:dyn_mass}). Thus, by considering the beam angular size, 
we estimate a molecular gas mass 5-$\sigma$ upper limit of $\sim 2.8 \times 10^{8}$\,$M_\odot$ to 
the three $\sim$kpc-scale gaseous clumps detected in the H$\alpha$ observation and reported by 
\citet{Swinbank2012b} for this galaxy.

\begin{figure}
\center
\includegraphics[width=0.8\columnwidth]{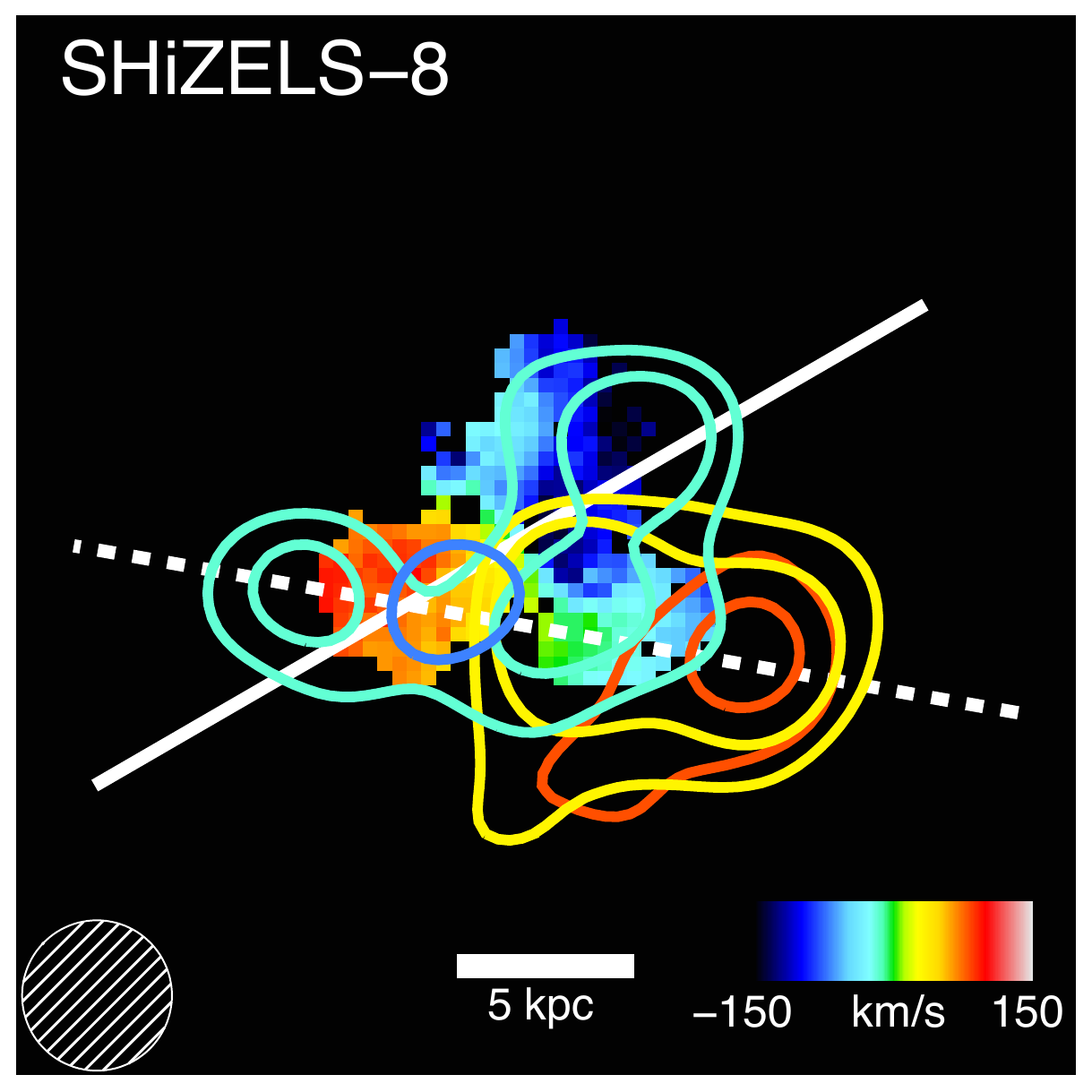}
\caption{ \label{fig:SHiZELS8_tapered}
H$\alpha$ velocity map at $\sim$\,kpc-scale for the SHiZELS-8 galaxy. The solid and dashed 
white lines represent the ionized and molecular gas major kinematic axis, respectively. The coloured 
contours represent the 3- and 5-$\sigma$ CO(2-1) emission from the 2000\,k$\lambda$ tapered datacube 
in four spectral channels ($\Delta v = 240$\,km\,s$^{-1}$). The synthesized beam size of this tapered 
observation ($\theta_{\rm BMAJ} = 0\farcs 50 \sim 4.3$\,kpc) is showed in the bottom-left corner. We note 
that the apparent shift between the two maps may be produced by the astrometry inaccuracies of the 
SINFONI observations. Nevertheless, we note that both observations suggest that the CO(2-1) and 
H$\alpha$ major kinematic axes are misaligned by $\sim$100-120\,deg, which is indicative of a kinematic
complex system.
}
\end{figure}

From the $0\farcs5$ smoothed map we obtain a velocity-integrated peak flux density r.m.s of 
2.5\,mJy\,km\,s$^{-1}$\,beam$^{-1}$. The lower image noise allows us to marginally 
detect the CO(2-1) emission in four spectral channels ($\Delta v = 240$\,km\,s$^{-1}$). 
We show the SHiZELS-8 marginally detection in Fig.~\ref{fig:SHiZELS8_tapered}. We clearly 
observe the CO emission line spatial and spectral shifts produced by the internal galactic 
dynamics. Thus, we estimate a rough major kinematic axis position angle (PA) of $\sim140$\,deg, 
with a peak-to-peak rotational velocity of $\sim$145\,km\,s$^{-1}$ (non-corrected by inclination).

\subsubsection{SHiZELS-8: a dynamically-perturbed system?}\label{sec:perturbedsystem}
Two pieces of evidence support the idea that SHiZELS-8 is a dynamically complex system. 
Firstly, the H$\alpha$ and CO dynamics show that both components rotate in the same direction but 
have position angles offset by $\sim 100-120$\,deg (see Fig.~\ref{fig:SHiZELS8_tapered}), which is in contrast 
to the negligible offset in the H$\alpha$ and CO dynamics seen in SHiZELS-19 (Fig.~\ref{fig:maps}). Second, 
our previous SINFONI observation shows a flat radial [N\,{\sc ii}]/H$\alpha$ metallicity gradient \citep{Swinbank2012a}.

We are possibly witnessing a massive reservoir of gas fuelling the star formation seen in 
H$\alpha$ in a similar way to that previously seen in more violent sub-millimetre galaxies (SMGs; \citealt{Tacconi2008}).
Indeed, the complex dynamics evidenced for the different ISM states might be mixing the
gas producing the flat metallicity gradient. We conclude that while SHiZELS-8 is a `typical' 
galaxy that resides in the upper range of the `main sequence' for star-forming galaxies, which follows the 
Kennicutt-Schmidt law (see Hughes et al. in prep.), it is probably experiencing torques that 
will eventually drive a flow of gas into the central regions. The SHiZELS-8 case demonstrates 
the wide variety of galaxy kinematics within the `main-sequence' \citep{Elbaz2018}. Given the 
impossibility to describe this source as a virialized rotating disk, in the remaining of this work 
we focus on the analysis of the SHiZELS-19 galaxy.

\subsection{The SHiZELS-19 galaxy}

We derive the two-dimensional intensity and kinematic maps for SHiZELS-19 by analysing 
the CO(2-1) emission line following the approach presented in \citet{Swinbank2012a}. Briefly, we 
spatially bin the ALMA data-cube up to a scale given by the synthesized beam and then we perform 
an emission line fitting approach using a $\chi^2$ minimization procedure (see \citealt{Swinbank2012a}, 
for more details). In each iteration a Gaussian profile is fitted in the frequency domain to 
estimate the intensity, velocity and velocity dispersion information (Fig.~\ref{fig:maps}). We highlight 
that for this galaxy, the H$\alpha$ emission line properties were derived and presented in an 
analogous manner in \citet{Molina2017}.

We show the CO(2-1) intensity, velocity and line-of-sight velocity dispersion maps for SHiZELS-19
in Fig.~\ref{fig:maps}, whilst the best-fitted kinematic parameters are listed in Table~\ref{tab:table3}. 
We observe a smooth CO(2-1) intensity map with no apparent clumpiness, which is 
consistent with the morphology observed in the H$\alpha$ intensity map \citep{Molina2017} and the 
\textit{HST} F160W-band (reft-frame optical) image. However, this galaxy presents an irregular morphology
in the \textit{HST} F814W-band map (rest-frame UV, Fig.~\ref{fig:fig1}). The discrepancy between the 
galaxy morphology seen in the \textit{HST} images suggests that the irregular morphology seen in the \textit{HST} 
F814W-band image may just reflect a complex dust distribution through the galactic disk (e.g. \citealt{Genzel2013}).

\subsubsection{Global Dynamical Properties}
\label{sec:galaxy_dyn}

\begin{figure*}
\includegraphics[width=0.41\columnwidth]{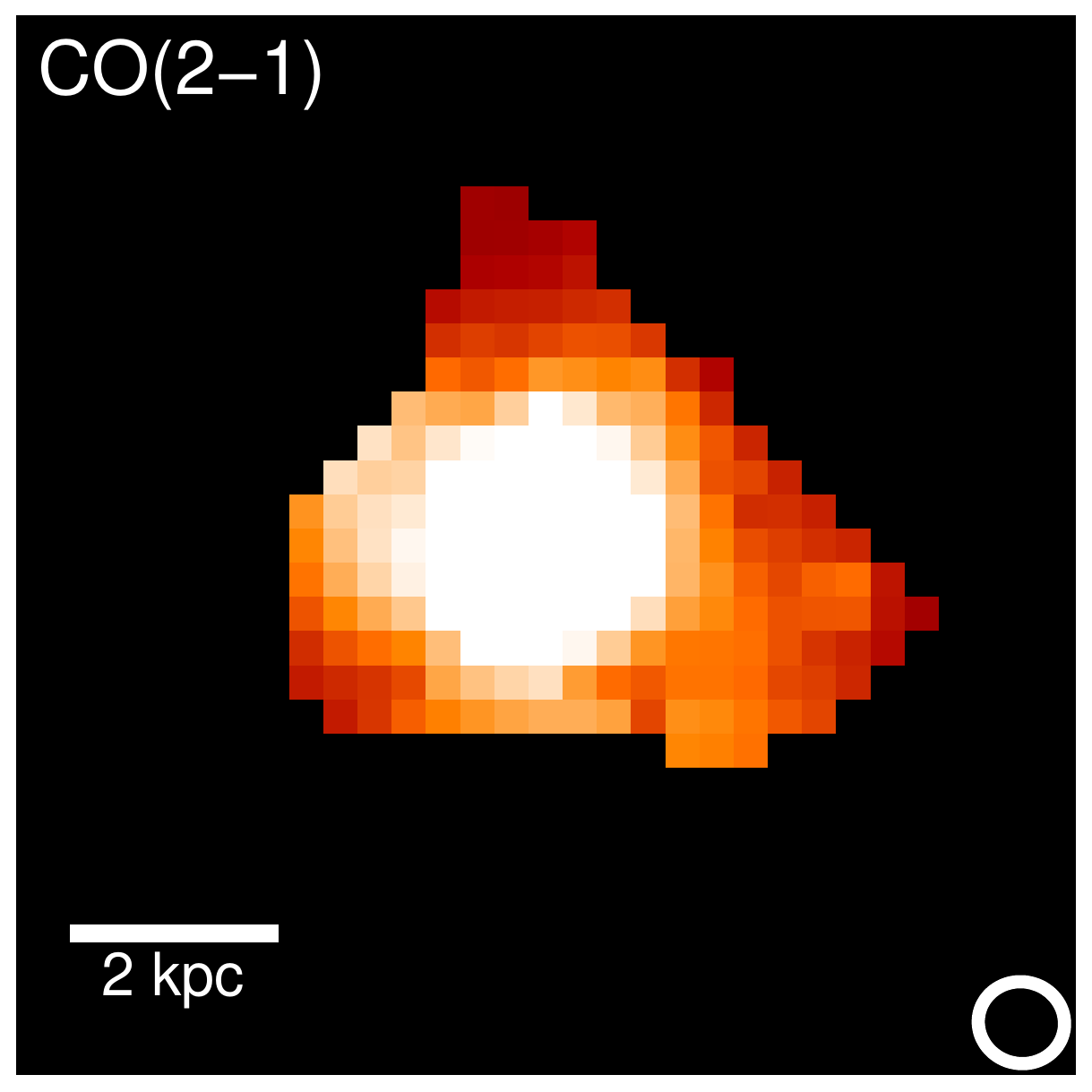}
\includegraphics[width=0.41\columnwidth]{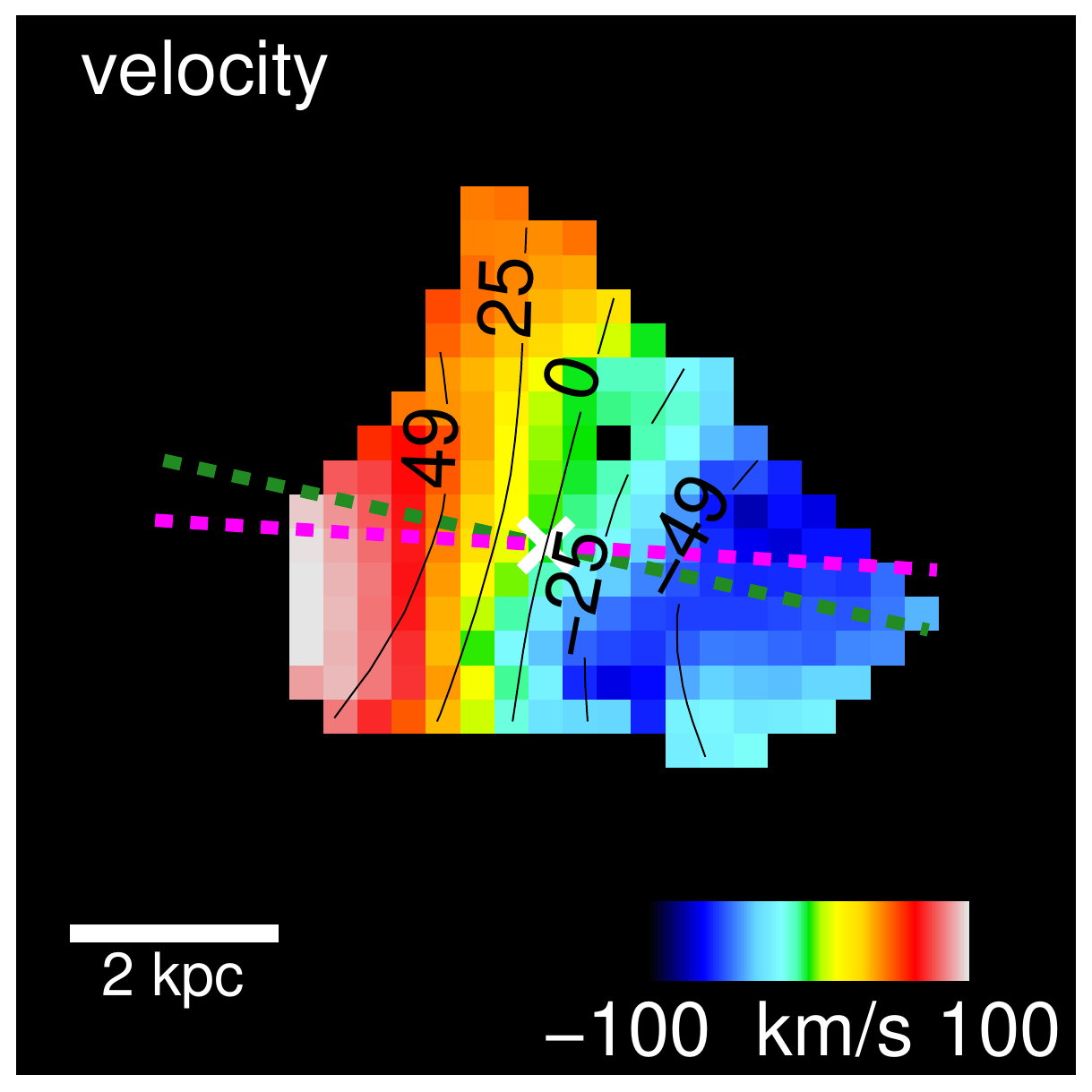}
\includegraphics[width=0.41\columnwidth]{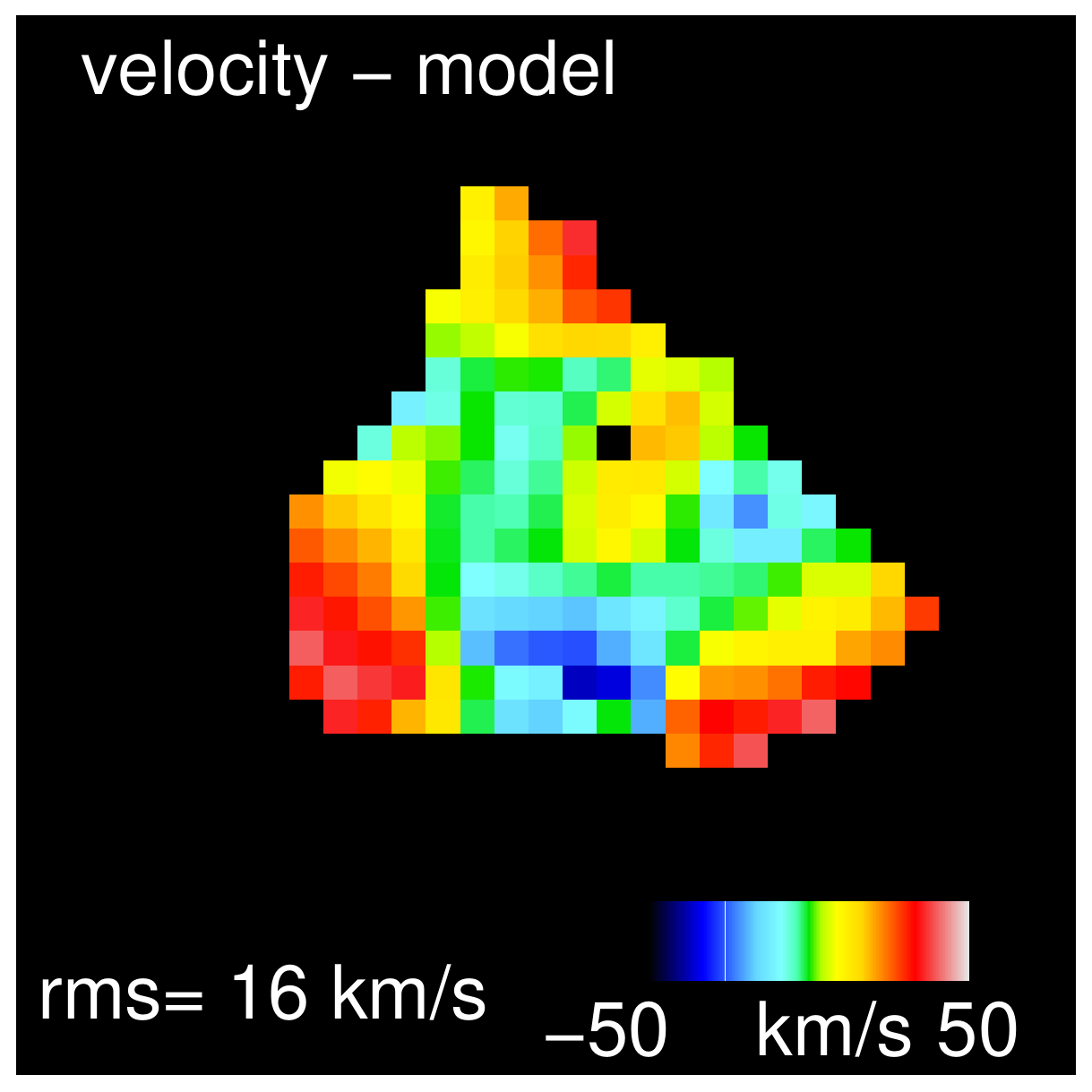}
\includegraphics[width=0.41\columnwidth]{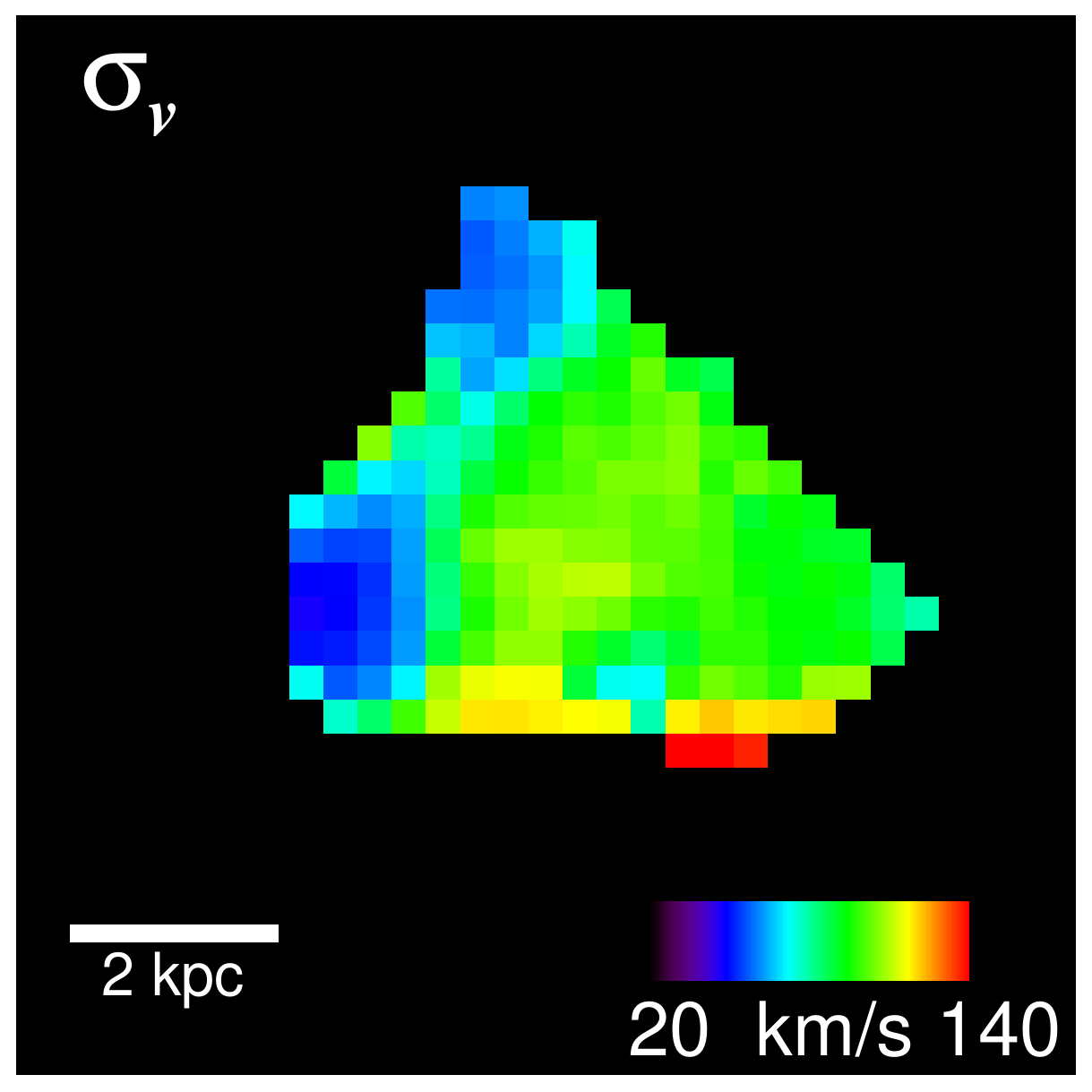}
\includegraphics[width=0.41\columnwidth]{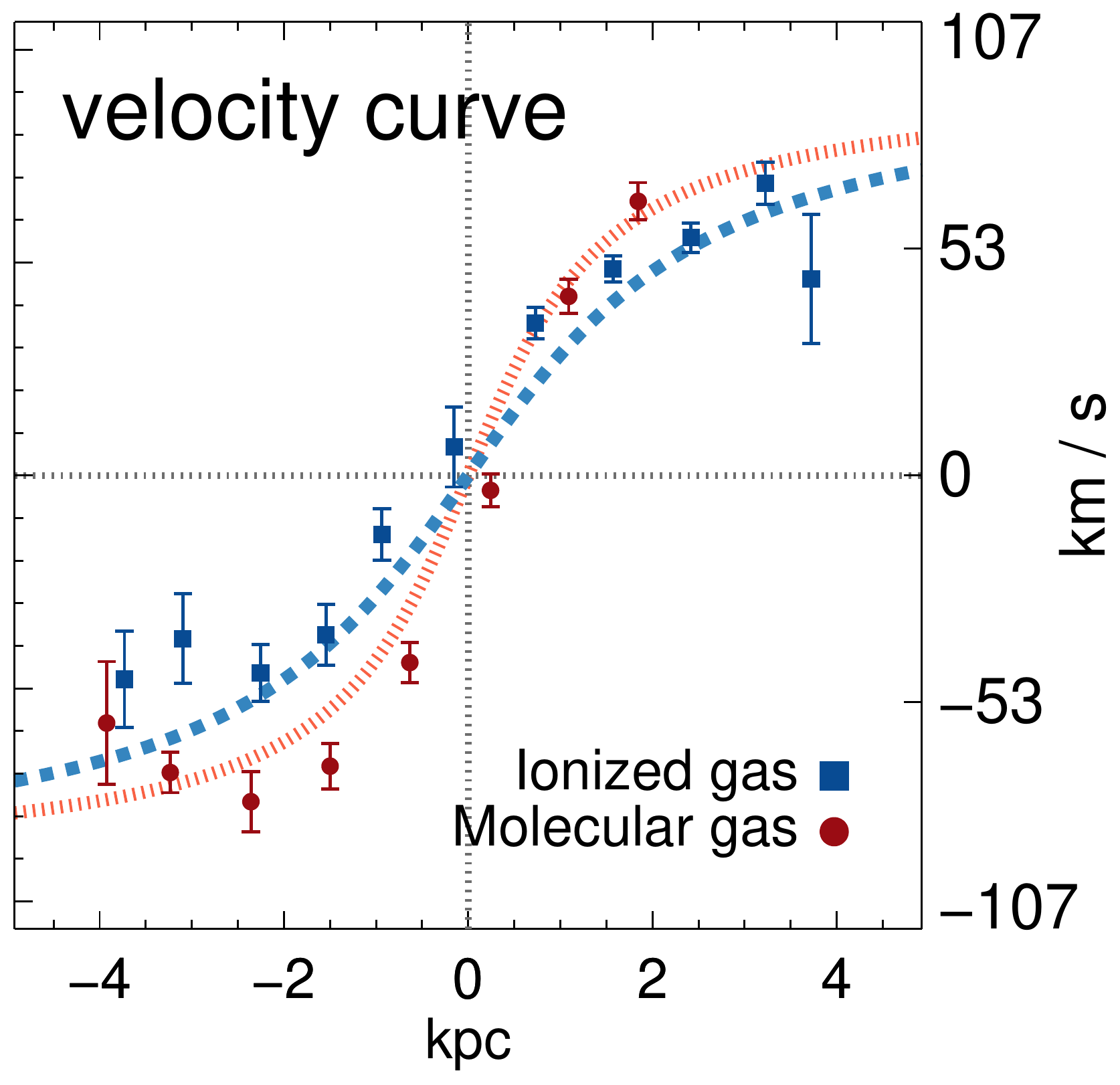}\\
\includegraphics[width=0.41\columnwidth]{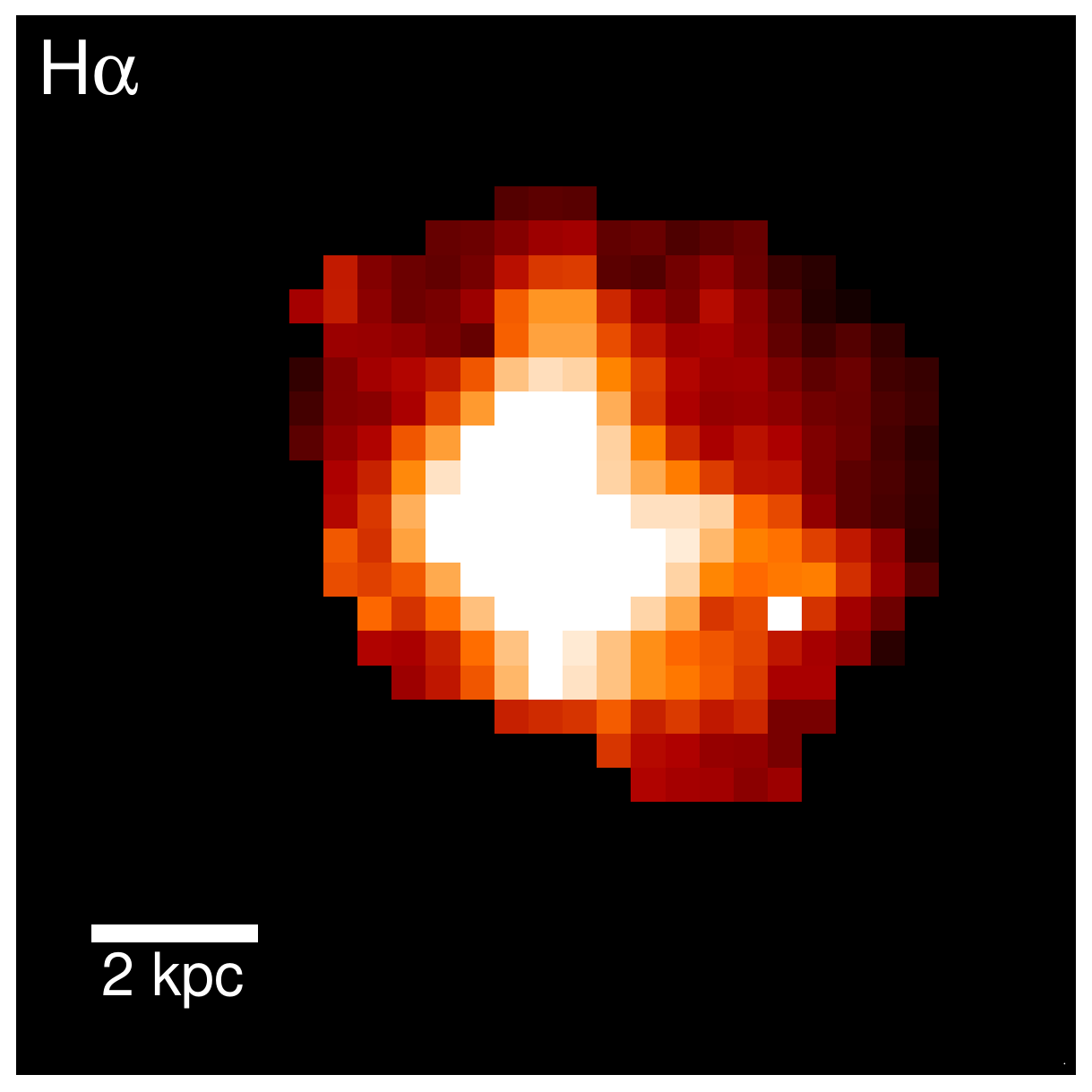}
\includegraphics[width=0.41\columnwidth]{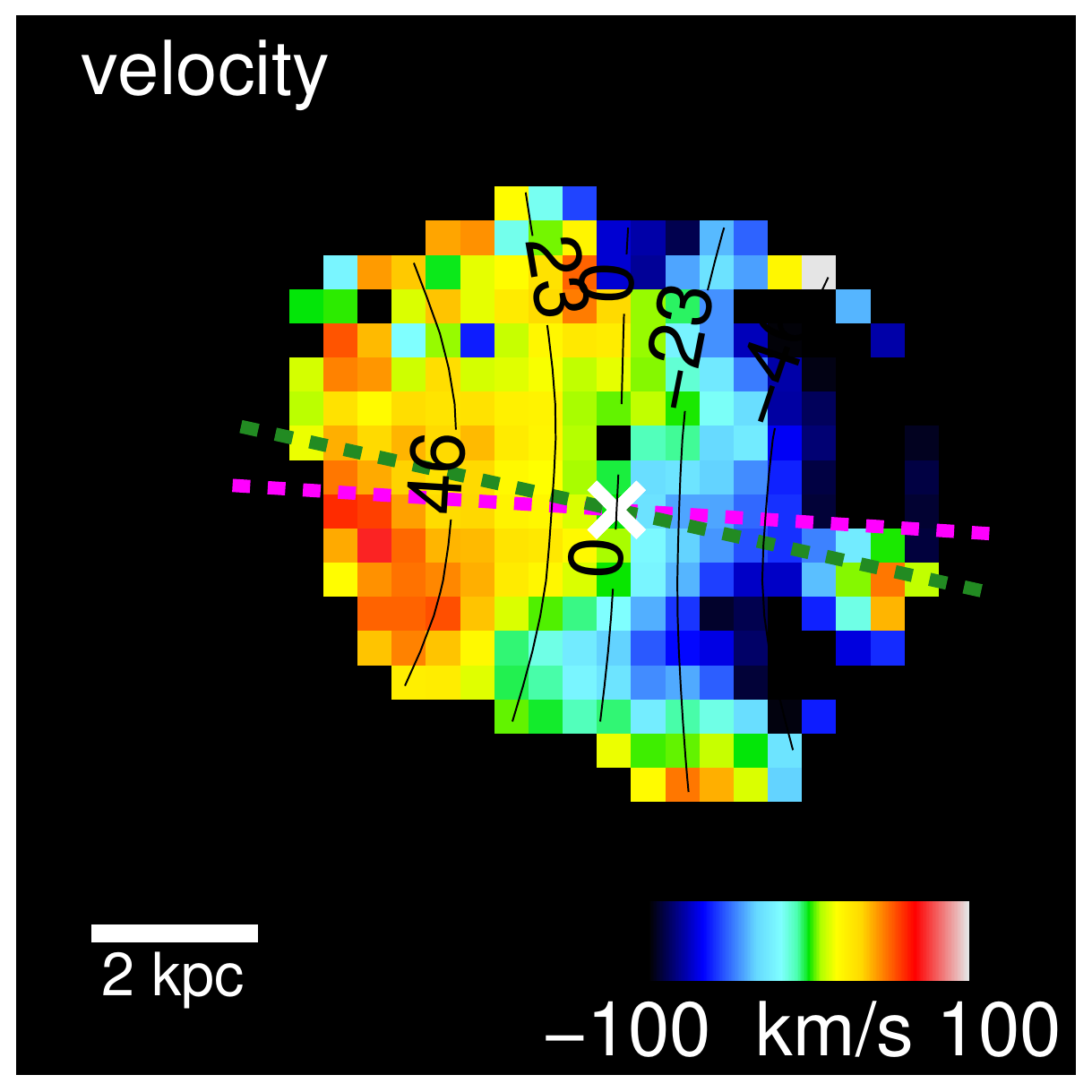}
\includegraphics[width=0.41\columnwidth]{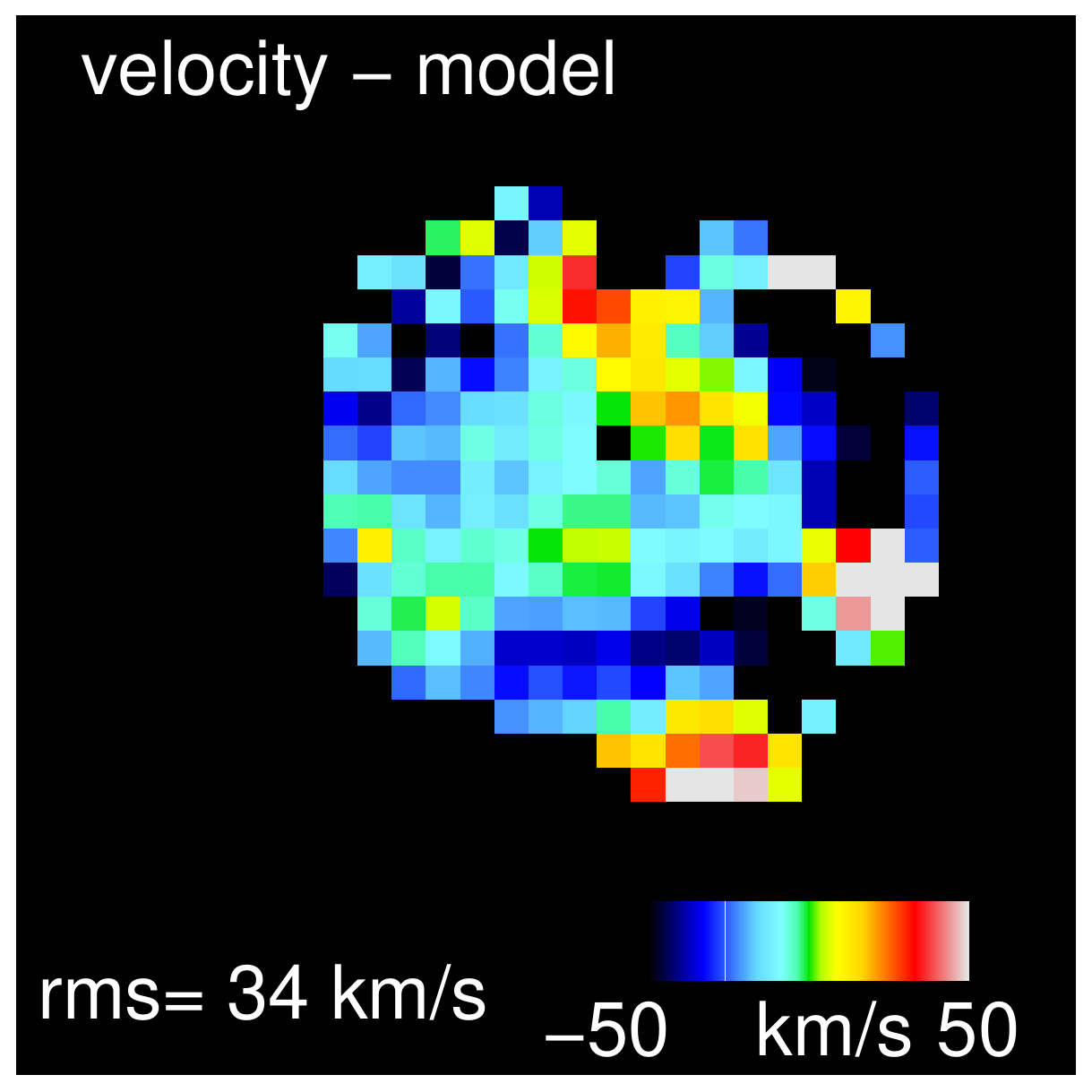}
\includegraphics[width=0.41\columnwidth]{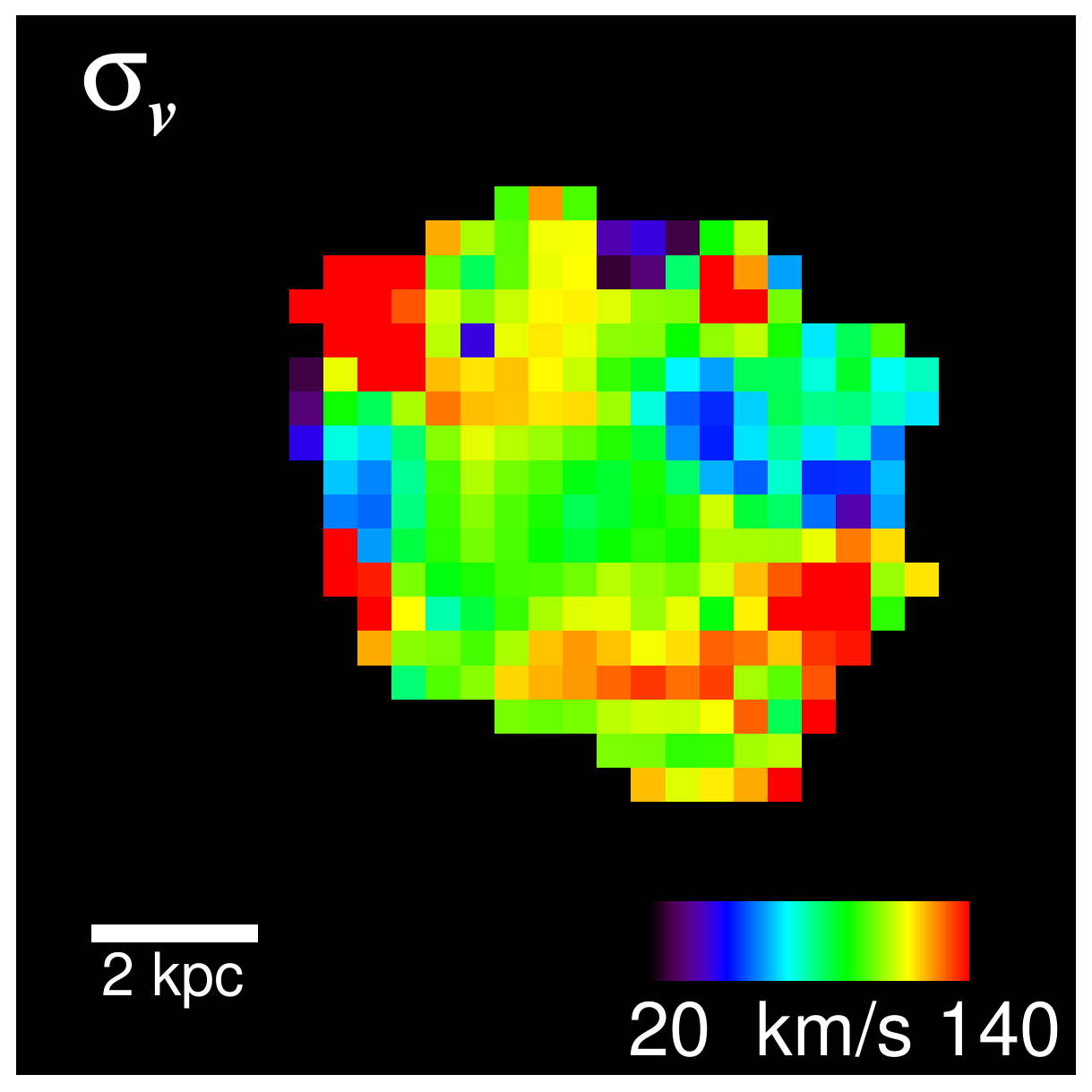}
\includegraphics[width=0.41\columnwidth]{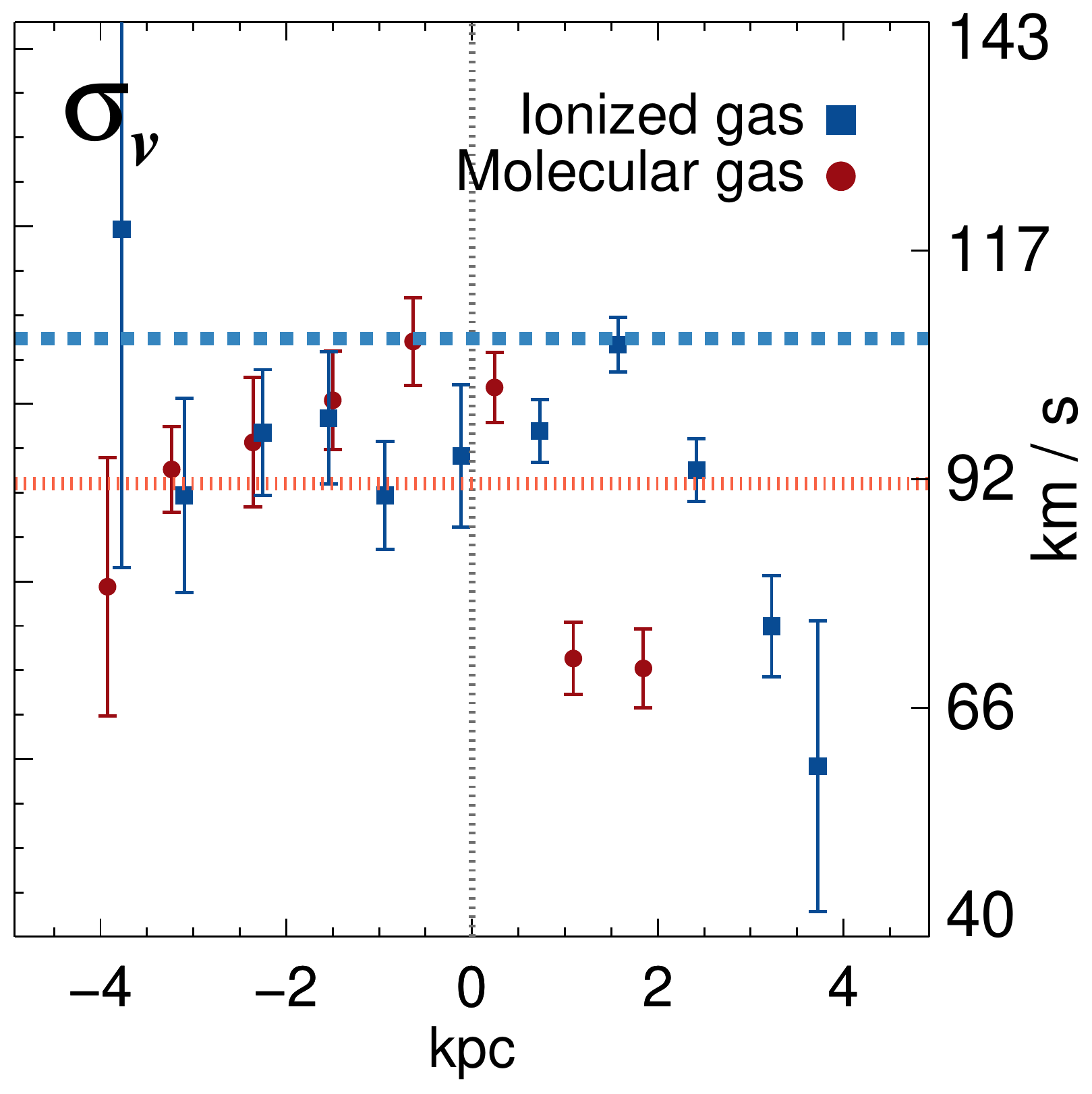}
\caption{ \label{fig:maps}
Intensity, velocity, residual and velocity dispersion maps (1st to 4th columns) for SHiZELS-19
obtained from CO(2-1) (top) and H$\alpha$ (bottom) emission lines. In the 5th column 
we show the one-dimensional rotational velocity (top) and velocity dispersion (bottom) profiles 
across each major kinematic axis for both observations. The spatial scale for each observation 
is showed in each moment map. The CO(2-1) intensity map also shows the synthesized 
beam size. The velocity maps have over-plotted the
kinematic centre and the velocity contours from their best-fit disk model. The green- and 
pink-dashed lines represent the molecular and ionized gas major kinematic axes, respectively.
The residual fields are constructed by subtracting the velocity disk model from the velocity
maps: the r.m.s. of these residuals are given in each panel. The velocity dispersion
maps are corrected for beam-smearing effects. The one-dimensional profiles are constructed 
by using the best-fit kinematic parameters and a slit width equal to half of the synthesized 
beam/PSF FWHM. In each one-dimensional profile, the error bars show the 1-$\sigma$
uncertainty and the vertical dashed grey line represents the best-fit dynamical centre. In 
the velocity profile panel, the red- and blue-dashed curves show the velocity curve extracted
from the beam-smeared CO and H$\alpha$ two-dimensional best-fit models, respectively. In 
the $\sigma_v$ one-dimensional profile panel, the red- and blue-dashed lines show the average 
galactic value (Table~\ref{tab:table3}) for the CO and H$\alpha$ observation, respectively.
}
\end{figure*}

{In order to characterise the dynamical properties of SHiZELS-19, we fit the 
two-dimensional velocity fields for the ionized and molecular gas jointly.}
We construct two-dimensional models with an input rotation curve following an arctan function 
[$V(r)=\frac{2}{\pi}V_{\rm asym}$arctan($r/r_{\rm t}$)], where $V_{\rm asym}$ is the asymptotic 
rotational velocity and $r_{\rm t}$ is the effective radius at which the rotation curve turns over 
\citep{Courteau1997}. We consider the `disk thickness' by modelling the galaxy as an oblate spheroid 
system with intrinsic minor-to-major axis ratio of 0.2, a value that seems appropriate for the high 
redshift galaxy population \citep{Law2012a}. As the CO and H$\alpha$ velocity fields 
are consistent (Fig.~\ref{fig:maps}), we also model both velocity fields by coupling the inclination
angle parameter. We do not attempt to lock the dynamical centres through RA$-$DEC coordinates 
as the SINFONI astrometry is not accurate enough to allow it, nevertheless we are assuming that the 
ionized and molecular gas ISM phases are co-planar. We also allow the possibility that their rotational 
motions can be out of phase, i.e. both ISM phases could have different kinematic PA. 

This modelling has eleven free parameters ($V_{\rm asym, H\alpha}$, 
$r_{\rm t, H\alpha}$, PA$_{\rm H\alpha}$, [x/y]$_{\rm H\alpha}$, 
$V_{\rm asym, CO}$, $r_{\rm t, CO}$, PA$_{\rm CO}$, [x/y]$_{\rm CO}$ 
and inclination angle; see Table~\ref{tab:table3}) and a genetic algorithm \citep{Charbonneau1995} 
is used to find the best-fit model (see \citealt{Swinbank2012a} for more details).
The total $\chi^2$ of the model is calculated as the sum of the $\chi^2$
obtained from each two-dimensional modelled map.

In \citet{Molina2017} the kinematic model for SHiZELS-19 was performed without any constraint 
on the inclination angle value. This adds an additional source of uncertainty as the inclination angle is
poorly constrained from the velocity field modelling alone \citep{Glazebrook2013}. In order to deal with this uncertainty,
we constrain the inclination angle by fitting a two-dimensional S\'ersic model \citep{Sersic1963} to the CO 
intensity map (moment 0) using \texttt{GALFIT} \citep{Peng2010}. We obtain an observed minor-to-major 
axis ratio of $\sim 0.90\pm 0.05$, which corresponds to an inclination angle value of $\sim26 \pm 6$\,deg. 
However, as \texttt{GALFIT} tends to underestimate the parameter errors, we consider a more conservative 
inclination angle uncertainty of $\pm$10\% in our fitting procedure \citep{Epinat2012}. 

We use the dynamical centres and position angles derived from the best-fit
dynamical models to extract the one-dimensional rotation curve and velocity 
dispersion profile across the major kinematic axes of the ionized and molecular 
gas. The extracted one-dimensional rotational curves and dispersion velocity 
profiles are presented in Fig.~\ref{fig:maps}.

We calculate the half-light radius ($r_{1/2}$) for each ISM component by following 
\citet{Molina2017}, and we define the rotational velocity for the ionized and molecular 
gas component ($V_{\rm rot, H\alpha}$, $V_{\rm rot, CO}$) as the inclination-corrected 
velocity observed at two times the H$\alpha$ and CO half-light radii, respectively 
(see Table~\ref{tab:table3}).

Even at the $\sim$kpc-scales achieved here, there is still a contribution to the derived 
line widths from the beam-smeared large-scale velocity motions across the galaxy 
\citep{Davies2011}. In order to correct for these effects, we calculate the velocity 
gradients ($\Delta V$/$\Delta R$) across the synthesized beam and Point Spread 
Function (PSF) in the CO and H$\alpha$ velocity field models, respectively. We 
subtract them linearly from the corresponding velocity dispersion map by following 
Eq.\,A1 from \citet{Stott2016}. However, by using this procedure, $\sim20$\% 
residuals are expected to remain, especially at the centres of each galaxy map 
where large velocity gradients are expected to be present \citep{Stott2016}. In order 
to minimize such effects, we define the global velocity dispersion for each gas phase
($\sigma_{v,{\rm CO}}$, $\sigma_{v,{\rm H\alpha}}$) as the median value taken from 
the pixels beyond the central galactic zone. This zone is defined as three times the 
size of the angular resolution of the map.

The best-fit kinematic maps and velocity residuals for the H$\alpha$ and CO derived 
maps are shown in Fig.~\ref{fig:maps}. The best-fit inclination, position angle 
and half-light radius values are given in Table~\ref{tab:table3}. The mean deviation
from the best-fit model (indicated by the typical r.m.s) is given in each residual map.

The molecular and ionized gas components show similar scale sizes $r_{\rm 1/2,H\alpha} / r_{\rm 1/2,CO} \approx 1.07 \pm 0.09$.
We stress that the CO and H$\alpha$ analyses are obtained from images created at matched spatial resolution (0$\farcs$15; 
corresponding to $\sim$kpc-scale at $z \sim 1.47$). Possible loss of the extended CO flux in the high-resolution 
observation may reduce the $r_{\rm 1/2,CO}$ value in our calculation.  Nevertheless, our estimation of the 
half-light radius for both ISM components are slightly smaller than the half-light radius value 
measured from the \textit{HST} F160W-band image $r_{\rm 1/2,HST-F160W} = 2.1 \pm 0.5$\,kpc \citep{Gillman2019}.

\begin{table}
\centering
	TABLE 3: KINEMATIC PROPERTIES\\
    	\caption{\label{tab:table3}
	  Best-fit kinematic parameters for SHiZELS-19 galaxy. `inc.'\ is the inclination angle 
	  defined by the angle between the line-of-sight (LOS) and the plane perpendicular 
	  to the galaxy disk (for a face-on galaxy, inc. = 0 deg.). The velocity dispersion 
	  and half-light radii values are corrected for `beam smearing' effects (see 
	  \S~\ref{sec:galaxy_dyn} for more details). The last row shows the reduced 
	  chi-squared ($\chi^2_\nu$) of the best-fit two-dimensional model.
      }
	\begin{tabular}{lcc} 
        \hline
        \hline
        ID & SHiZELS-19 \\
        \hline
        PA$_{\rm H\alpha}$ (deg) & 176$\pm$18 \\
        $\sigma_{v,{\rm H\alpha}}$ (km\,s$^{-1}$) & 107$\pm$13 \\
        $V_{\rm rot,H\alpha}$ (km\,s$^{-1}$) & 106$\pm$9 \\
        $r_{1/2,{\rm H\alpha}}$ (kpc) & 1.80$\pm$0.16 
        \vspace{1.5mm}\\
        PA$_{\rm CO}$ (deg) & $167\pm14$ \\
        $\sigma_{v,{\rm CO}}$ (km\,s$^{-1}$) & 91$\pm$6\\
        $V_{\rm rot,CO}$ (km\,s$^{-1}$) & 121$\pm$10 \\
        $r_{1/2,{\rm CO}}$ (kpc) & 1.68$\pm$0.03         
        \vspace{1.5mm} \\
        inc. (deg) & 27.5$\pm$0.6\\
        $\chi^2_\nu$ & 3.51\\ 
        \hline
	\end{tabular}
\end{table}

The CO(2-1) velocity map shows a clear rotational pattern, roughly matching
the rotational motions traced by the ionized gas component.
From the two-dimensional modelling, we find that the kinematic position angles 
agree ($\Delta$PA\,$\equiv$\,PA$_{\rm H\alpha}-\,$PA$_{\rm CO} = 9 \pm 23$\,deg) 
within the 1-$\sigma$ error range. The velocity curves roughly agree, except in the
blueshifted zone where the CO traced rotation curve drops to lower velocity 
values. However, we note that the ionized gas velocity map is noisier 
than the molecular gas velocity map, especially in the galaxy outskirts. This may be 
partly produced by OH line features present in the $H-$band spectra, whilst the 
ALMA observation is free from sky-line residuals. We find that, the ionized
gas component shows a slightly lower rotational velocity value when compared to that
from the molecular gas observations ($V_{\rm rot,H\alpha} / V_{\rm rot,CO} \approx 0.88 \pm 0.10$).
This might be due to differences in the spatial distribution between the two ISM components.

In terms of velocity dispersion, the CO observation shows a slightly lower average 
velocity dispersion value than the mean value observed from the ionized gas 
component ($\sigma_{v,{\rm H\alpha}}$/$\sigma_{v,{\rm CO}} \approx 1.18 \pm 0.16$).
The difference between the ALMA (25\,km\,s$^{-1}$) and $H-$band SINFONI 
(50\,km\,s$^{-1}$) spectral resolutions should not produce such differences as the 
intrinsic CO and H$\alpha$ line-widths are significantly broader.
The high $\sigma_v$ values observed at the outskirts of the H$\alpha$ velocity 
dispersion map may increase the ionized gas average value. In a similar way as the 
comparison between the velocity maps, the ionized gas velocity dispersion map is noisier 
than the molecular gas map at larger radii. By considering all the pixels 
in the mean $\sigma_v$ estimation, we obtain an average $\sigma_{v,{\rm H\alpha}}$ value of 
$91 \pm 13$\,km\,s$^{-1}$ \citep{Molina2017}, in agreement with the measured $\sigma_{v,{\rm CO}}$
value (Table~\ref{tab:table3}). 
Thus, we suggest that both ISM tracers show similar supersonic turbulence values.
 
We derive rotational velocity to dispersion velocity ratio ($V_{\rm rot}/\sigma_v$) 
values of 0.99$\pm$0.14 and 1.33$\pm$0.14 for the ionized and molecular gas ISM 
phases, respectively. This suggests that the disordered motions of both ISM phases
are playing an important role in the galactic support against self-gravity \citep{Burkert2010}.

\begin{figure}
\includegraphics[width=\columnwidth]{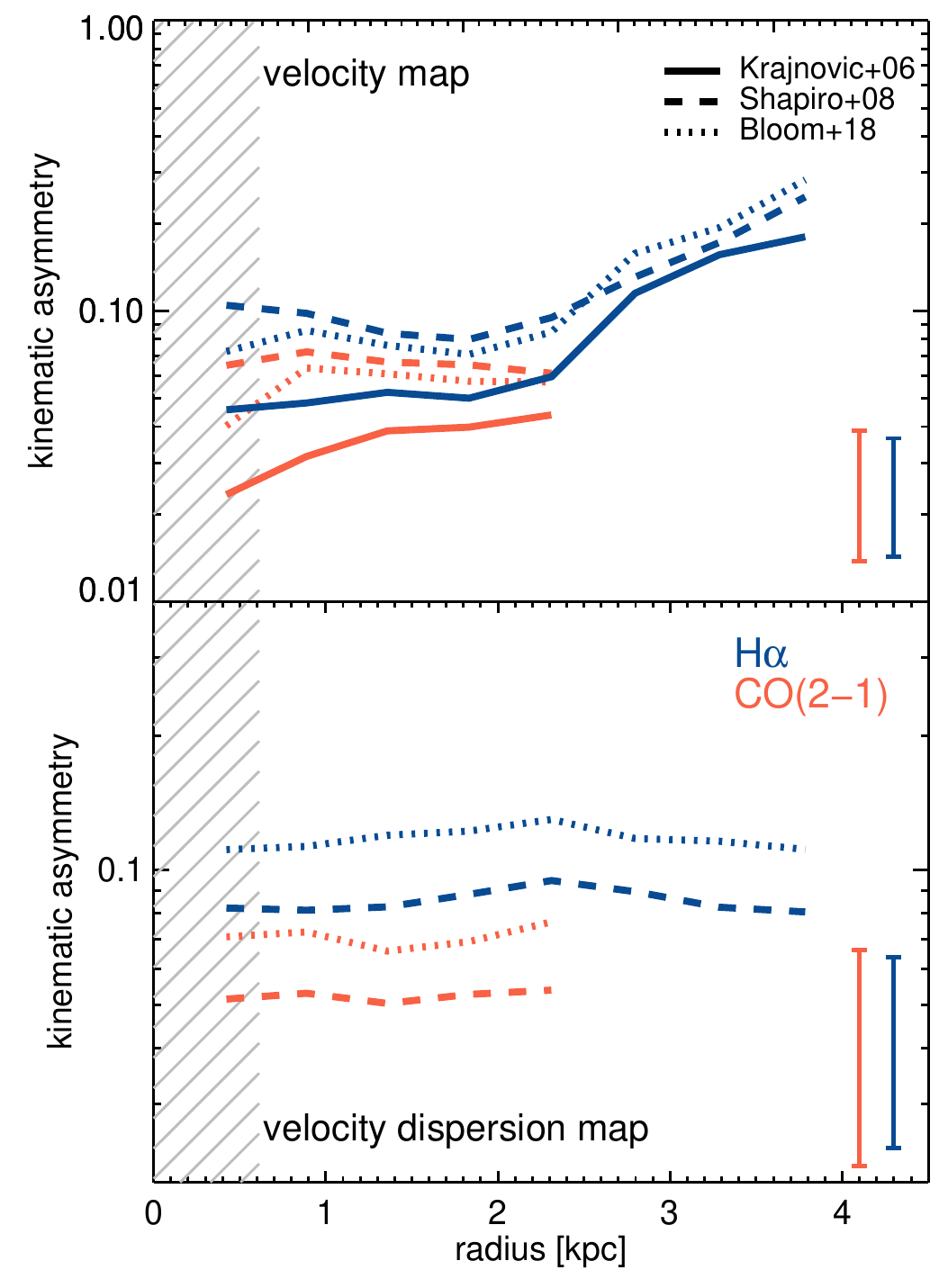}
\caption{\label{fig:asymmetry}
Kinematic asymmetry radial profiles measured from the SHiZELS-19 velocity (\textit{top}) and velocity dispersion 
(\textit{bottom}) maps. We plot the CO(2-1) and H$\alpha$ observations. 
The line, dashed-line and dotted-line represent the kinematic asymmetry estimators 
presented in \citet{Krajnovic2006,Shapiro2008} and \citet{Bloom2018} for each map 
(see \S~\ref{sec:kinematic_deviations} for more details). The colour-coded error bars show the median 
1-$\sigma$ uncertainties in each panel. The grey-dashed area represents the resolution element 
radial extent. Despite of the estimator used, the ionized gas two-dimensional 
maps tend to show slightly higher kinematic deviations from the ideal rotating disk case than the 
molecular gas kinematic maps. Although, the measurements agree within 1-$\sigma$ error range.}
\end{figure}

\subsubsection{Kinematic Asymmetry Characterization}
\label{sec:kinematic_deviations}

In order to obtain a detailed characterization of the ionized and molecular gas kinematics, we 
quantify the kinematic deviations from the ideal rotating disk case by performing a 
`\textit{kinemetry}' analysis \citep{Krajnovic2006}. Briefly, \textit{kinemetry} proceeds to analyse 
the two-dimensional kinematic maps using azimuthal profiles in an outward series of best fitting 
tilted rings. The kinematic profile as a function of angle is then expanded harmonically, which is 
equivalent to a Fourier transformation which has coefficients $k_{n,v}$ and $k_{n,\sigma}$ at 
each tilted ring for the velocity and velocity dispersion maps, respectively. In the velocity map, 
the first order decomposition `$k_{1,v}$' is equivalent to the rotational velocity value, and therefore, 
the ideal rotating disk case is simply described by the cosine law along the tilted rings 
($V(\theta) = k_{1,v}  \cos (\theta)$). The high-order terms describe the kinematic anomalies with 
respect to the ideal rotating disk case (see \citealt{Krajnovic2006} for more details).
We note that \textit{kinemetry} stops the radial fitting when there are less than 75\% of the pixels 
sampled along the best-fit tilted ring \citep{Krajnovic2006}.

We restrict the inclination and position angles within the 1-$\sigma$ error 
range given by our best-fit two-dimensional model. The $k_{n,v}$ and $k_{n,\sigma}$ 
errors are derived by bootstrapping via Monte-Carlo simulations the errors in measured 
velocities, velocity dispersions, and estimated dynamical parameters.

We quantify the kinematic deviations from the ideal disk case by computing three 
different estimators used in the literature: (1) the $k_{5,v}/k_{1,v}$ ratio \citep{Krajnovic2006}; 
(2) the $(k_{2,v} + k_{3,v} + k_{4,v} + k_{5,v}) / 4k_{1,v}$ and 
$(k_{1,\sigma} + k_{2,\sigma} + k_{3,\sigma} + k_{4,\sigma} + k_{5,\sigma}) / 5k_1,v$ 
fractions \citep{Shapiro2008}; and (3) the $(k_{3,v} + k_{5,v}) / 2k_{1,v}$ and 
$(k_{2,\sigma} + k_{4,\sigma}) / 2k_{1,v}$ ratios \citep{Bloom2018}. The first case is the 
traditional dimensionless ratio that describes the kinematic asymmetries just in the velocity 
map. It does not consider the low-order coefficients as these are used by `\textit{kinemetry}'
to find the best-fitted tilted rings at a given radius \citep{Krajnovic2006}. The second case 
was defined to classify galaxy mergers which tend to present extremely disturbed 
kinematic fields \citep{Shapiro2008}. The third case consists on a slight modification to the 
second case as it takes into account that in moderately disturbed systems, the even/odd moments 
contribution measured from the velocity/velocity dispersion maps are negligible \citep{Bloom2018}.

In Fig.~\ref{fig:asymmetry} we show the different estimators of the kinematic deviations for the CO 
and H$\alpha$ velocity and velocity dispersion maps as a function of the de-projected radius. We 
note that the shorter CO radial profiles compared to the H$\alpha$ radial profiles are produced by 
the stop of the `kinemetry' procedure at shorter radius due to the lack of roundness of the 
CO two-dimensional maps derived from our observations.

In the case of the velocity map, the $k_{5,v}/k_{1,v}$ \citep{Krajnovic2006} ratio gives lower values along 
the galactic disk compared with the other two estimators. We obtain an average $k_{5,v}/k_{1,v}$ ratio of 
0.04$\pm$0.01 and 0.09$\pm$0.05 for the CO and H$\alpha$ velocity map respectively. This difference is 
mainly produced by the higher $k_{5,v}/k_{1,v}$ values found in the H$\alpha$ velocity map at longer 
radius ($\gtrsim 2$\,kpc). This gradient suggests that SHiZELS-19 suffered a merger event in the past as the 
outer regions retain better the kinematic perturbations by remaining out of equilibrium while the central region 
tends to relax faster to a disk-like system \citep{Kronberger2007}.

If we follow the kinematic classification performed to the ATLAS$^{\rm 3D}$ \citep{Krajnovic2011} and 
SAMI local galaxy surveys \citep{vandenSande2017} and we consider their $k_{5,v}/k_{1,v} = 0.04$ limit value
to classify systems as regular rotators, this would imply that SHiZELS-19 corresponds to a `non-regular' rotator, i.e, 
the velocity field presents significant kinematic deviations that make it not well-described by the cosine law.

In the case of the velocity dispersion map, we found that the \citet{Bloom2018}'s estimator is higher than the
\citet{Shapiro2008}'s estimator at all radii. The additional $k_{n,\sigma}$ coefficients considered in the latter case
contribute little to the kinematic asymmetry estimator. This may also suggests that SHiZELS-19 is a moderate 
disturbed system. We also note that, as a difference from the velocity map, the kinematic asymmetries in the 
CO and H$\alpha$ velocity dispersion map tend to be nearly constant along the galactic disk. The kinematic 
deviations measured from the CO velocity dispersion map tend to be lower than the ones measured from the 
H$\alpha$ velocity map, however, they agree between the 1-$\sigma$ error range.

The rough agreement between the molecular and ionized gas kinematics suggests that, at $\sim$kpc-scales,  
both phases of the ISM are tracing the galactic dynamics instead of peculiar kinematics (e.g. gas 
inflows/outflows). This is in agreement with previous studies of massive galaxies 
(at $\sim 0.4-2.4 \times 10^{11}$\,$M_\odot$) at similar redshift (e.g. \citealt{Ubler2018,Calistro2018}). 

\subsection{Dynamical Mass \& Dark Matter content}
\label{sec:dyn_mass}

The dynamical mass estimate is a useful tool that allows us to measure the
total galactic mass enclosed as a function of radius. It provides a simple way to
probe the existence of dark matter haloes (e.g. \citealt{Gnerucci2011}) or to
constrain the CO-to-H$_2$ conversion factor (e.g. \citealt{Motta2018, Calistro2018}).

By measuring the global kinematics of a galaxy, the dynamical mass can be 
easily estimated from the rotational velocity (e.g. \citealt{Genzel2011})
considering a thin-disk dynamical mass approximation ($M_{\rm dyn,thin}$). 
On the other hand, if the supersonic turbulence across the galactic disk is 
comparable to the ordered motions amplitude, then, an additional pressure 
gradient support contribution against self-gravity has to be considered. In this 
limit, the galactic disk height is not negligible and a thick-disk approximation 
($M_{\rm dyn,thick}$) should be considered \citep{Burkert2010}.

\begin{figure}
\includegraphics[width=0.49\columnwidth]{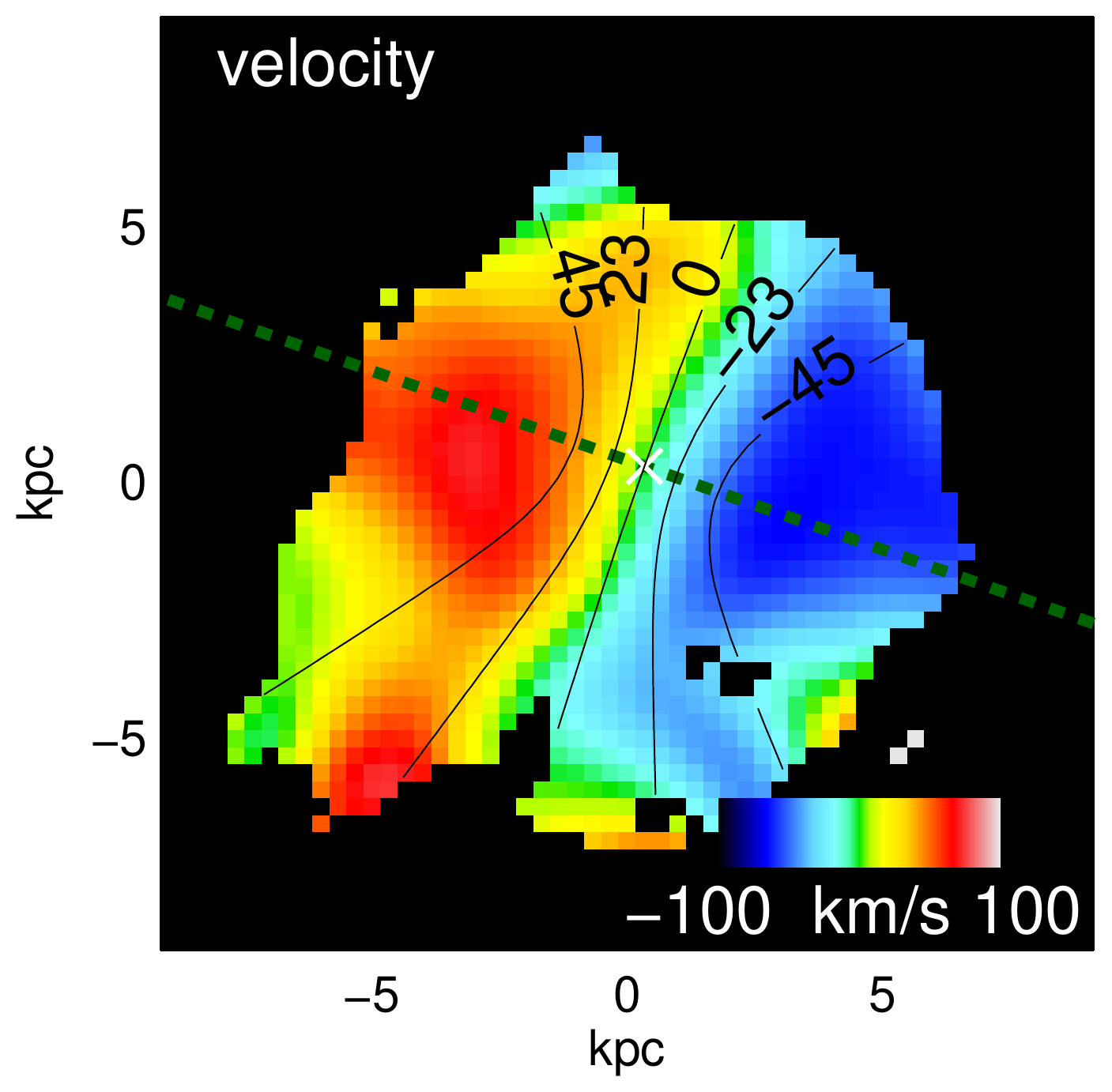}
\includegraphics[width=0.505\columnwidth]{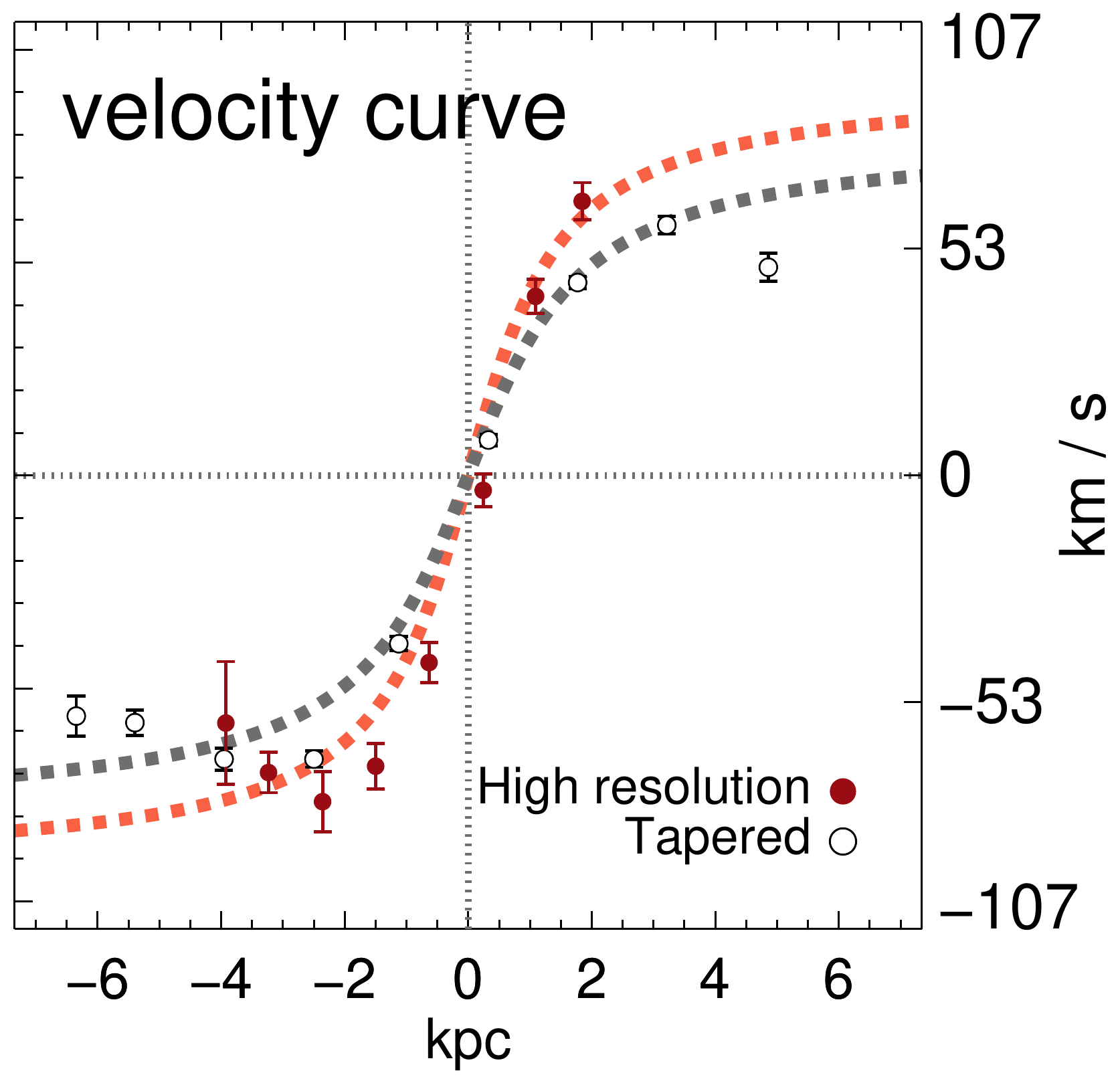}
\caption{ \label{fig:COS30_tapered}
Velocity map, and velocity profile for the CO(2-1) 2000\,k$\lambda$ tapered datacube for SHiZELS-19.
We use the same colour-coding presented in Fig.~\ref{fig:maps}. In this case, the 
one-dimensional velocity profile is constructed by using the best-fit kinematic parameters 
for the tapered datacube and a slit width equal to half of the tapered synthesized beam 
FWHM. We also overplot the data taken from the $\sim$kpc-scale high-resolution 
observations. The tapered rotation curve extends up to $\approx$6\,kpc.
}
\end{figure}

We calculate the dynamical mass for the SHiZELS-19 galaxy by using the kinematic 
information from our CO observations as its velocity map shows lower kinematic asymmetry 
amplitudes compared to the H$\alpha$ velocity map (\S~\ref{sec:kinematic_deviations}). Since 
the CO $\sim$kpc-scale observations are more sensitive to the denser and compact emission, 
we use a tapered version (2000\,k$\lambda$) of the ALMA observations that allows us to trace
the diffuse and more extended CO emission (at 0$\farcs$29 resolution). This allows us to observe 
a rotation curve up to a radial distance of $\approx$6\,kpc or $\sim$3.5 times the CO half-light 
radius (Fig.~\ref{fig:COS30_tapered}). 

Taking into account that the S\'ersic index derived from the \textit{HST} image is consistent with 
unity for this galaxy \citep{Gillman2019}, we assume an exponential disk surface density profile. This implies 
that, in terms of the disk scale length ($r_d$), we observe the rotation curve up to $\approx 6 r_d$ 
($r_{1/2} \approx 1.67 r_d$ for an exponential disk).

By using the inclination-corrected rotational velocity value derived from the tapered 
rotation curve at radius of $\approx$6\,kpc ($V_{\rm rot,tap} = 112 \pm 6$\,km\,s$^{-1}$), 
we would obtain a total enclosed mass of $M_{\rm dyn,thin}(r \lesssim 6$\,kpc$) = (1.75 \pm 0.19) \times 10^{10}$\,$M_\odot$ 
assuming a thin disk approximation. This dynamical mass estimate is lower 
but consistent within 1-$\sigma$ range with the estimated stellar mass for this galaxy scaled 
at the same radius ($M_\star(r \lesssim 6\,{\rm kpc}) \approx 0.98 M_\star \approx (1.96 \pm 0.90) \times 10^{10}$\,$M_\odot$). 
However, this `thin disk' dynamical mass value would suggest that this galaxy has almost no 
gaseous mass content, evidencing an apparent discrepancy with our CO and H$\alpha$ 
emission line measurements. On the other hand, as the $V_{\rm rot}/\sigma_v$ ratio is consistent 
with unity for both ISM components, this suggests that the $M_{\rm dyn,thin}$ quantity may be 
underestimating the total mass of this galaxy. Additional support against self-gravity needs to be 
considered.

We follow the analysis by \citet{Burkert2010}, and we consider a possible additional 
pressure support by calculating the dynamical mass in the thick-disk approximation.
In the `thick-disk' dynamical mass modelling, the radial pressure gradient 
term in the hydrostatic equation can not be neglected and it is parametrized 
by the galactic velocity dispersion and the mass density distributions. This approximation 
further assumes that $\sigma_{v}$ is independent of the galaxy disk radius and height. 
We use \citet{Burkert2010}'s Eq.11 with $\sigma_{v,{\rm CO}}$ and $r_{1/2,{\rm CO}}$ as 
input values and we obtain $M_{\rm dyn,thick} = (1.59\pm 0.19) \times 10^{11}$\,$M_\odot$. 
This dynamical mass value is $\sim 8$ times higher than $M_\star$, erasing any discrepancy 
between both quantities, but allowing the possibility of a non-negligible amount of dark matter 
content in this galaxy.

In order to test this, we calculate the dark matter fraction by comparing the total mass budget 
from our dynamical analysis with the luminosity-based total mass content. We 
consider the total $M_\star$ value estimated for the SHiZELS-19 galaxy as its difference with 
the scaled value at 6\,kpc ($M_\star - M_\star(r \lesssim 6\,{\rm kpc}) \approx 0.02 M_\star$) is 
negligible compared to the stellar mass uncertainty (see Table~\ref{tab:table2}). Therefore,
by considering the $M_\star$, $M_{\rm H_2}$ and $M_{\rm dyn}$ quantities, we define the dark 
matter fraction as,

\begin{equation}
\label{eq:f_dm}
f_{\rm DM} \equiv 1- \frac{M_\star + M_{\rm H_2}}{M_{\rm dyn,thick}} = 1 - \frac{\alpha_{\rm CO} L'_{\rm CO} + M_\star}{M_{\rm dyn,thick}}.
\end{equation}

\noindent where the molecular mass content is estimated via the CO luminosity 
($M_{\rm H_2}=\alpha_{\rm CO} L'_{\rm CO}$). However, this mass sum approach 
needs additional information about the CO-to-H$_2$ conversion factor in order to 
overcome the degeneracy between $\alpha_{\rm CO}$ and $f_{\rm DM}$. We also
note that strong dependence on the assumptions behind $M_\star$, $M_{\rm dyn}$ 
and $L'_{\rm CO}$ may also affect the result from Eq.~\ref{eq:f_dm}.

Thus, in order to properly consider the $M_\star$, $M_{\rm dyn}$ and $L'_{\rm CO}$ 
uncertainties and the degeneracy between $\alpha_{\rm CO}$ and $f_{\rm DM}$, we 
reproduce the parameter space built up in Eq.~\ref{eq:f_dm} by applying an MCMC 
technique following \citet{Calistro2018}. Briefly, based on the likelihood of the measured 
$L'_{\rm CO}$, $M_\star$ and $M_{\rm dyn}$ values, we sample the posterior 
probability density function (posterior PDF) for $\alpha_{\rm CO}$ and $f_{\rm DM}$ 
parameters using the \textsc{emcee} algorithm \citep{Foreman2013}.

We note that SED fitting techniques based on unresolved flux observations may 
lead to the underestimation of the galactic stellar mass values \citep{SS2018}. Thus, 
we consider an additional case in which we assume that the stellar mass content is 
being underestimated by a factor of two. This is likely to be an extreme case as 
suggested by \citet{SS2018} for galaxies with similar sSFR.
 
\begin{figure}
\includegraphics[width=\columnwidth]{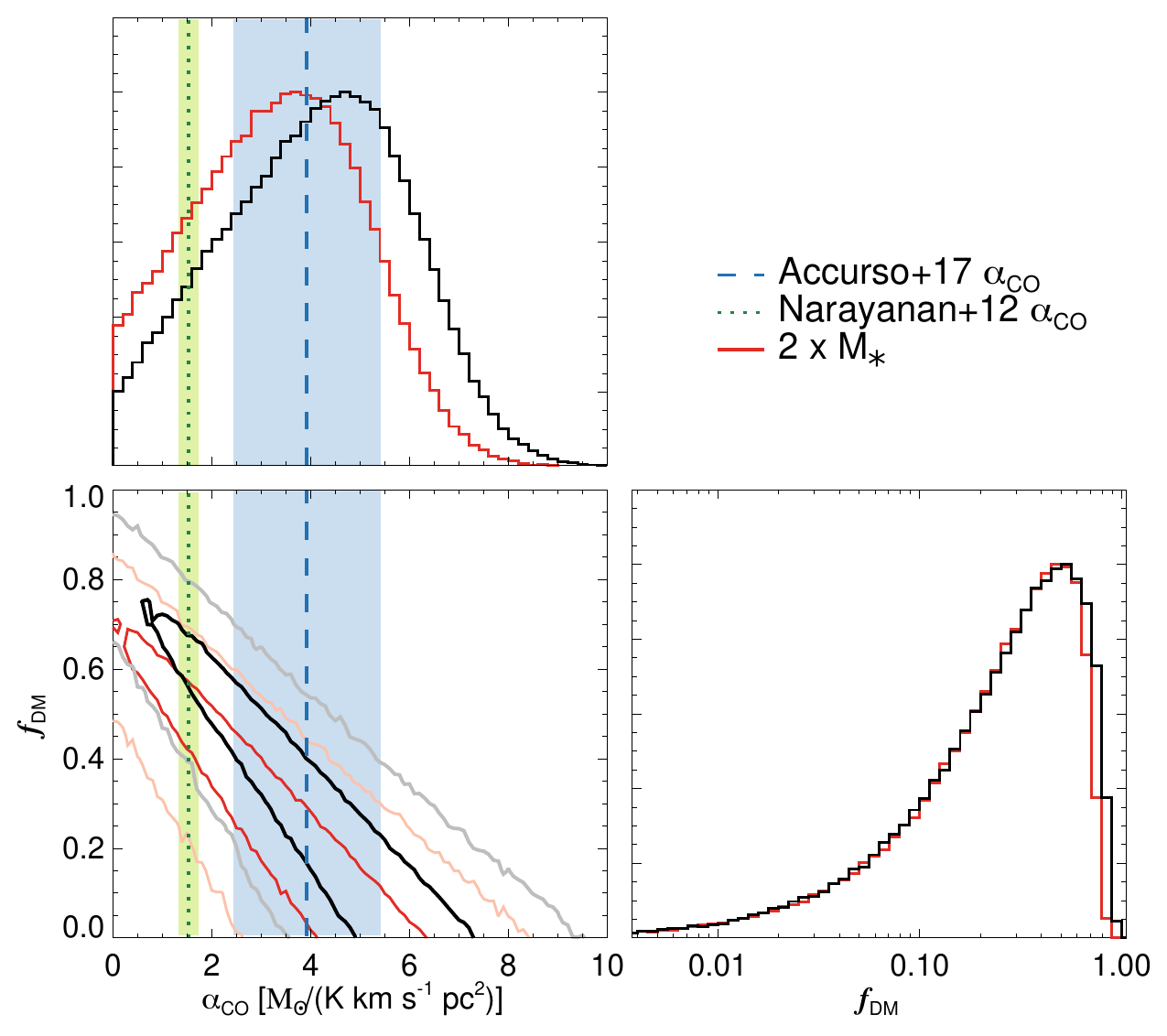}
\caption{ \label{fig:distributions}
One- and two- dimensional posterior PDFs of the $f_{\rm DM}$ and $\alpha_{\rm CO}$ parameters for SHiZELS-19.
The one-dimensional PDFs are represented by the black line in the \textit{top-left} and \textit{bottom-right} 
panels. The red line represents the inference assuming 2$\times$ the stellar-mass value derived from the spatially 
unresolved SED fitting, thus we consider a possible underestimation of $M_\star$ as suggested from spatially resolved 
studies \citep{SS2018}. In \textit{bottom-left} panel we show the two-dimensional PDFs, i.e., the covariance between 
both parameters. The black and grey lines show the 1- and 3-$\sigma$ regions of the PDF derived by using the 
stellar mas value obtained from the spatially-unresolved SED fitting. The red and orange lines show the 1- and 
3-$\sigma$ regions by assuming a stellar mass correction factor of two. In the \textit{bottom-} and \textit{top-left}
panels, the dashed and dotted lines show the CO-to-H$_2$ conversion factors calculated by following 
the \citet{Accurso2017} and \citet{Narayanan2012} parametrizations, respectively. The blue- and green- shaded 
regions show the 1-$\sigma$ uncertainties for each parametrization.
}
\end{figure}

In Fig.\ref{fig:distributions} we show the one- and two- dimensional posterior PDFs of the 
$\alpha_{\rm CO}$ and $f_{\rm DM}$ parameters. We also show the CO-to-H$_2$
conversion factor values suggested by following \citet{Accurso2017} and 
\citet{Narayanan2012}. From the two-dimensional posterior PDF we observe the 
strong degeneracy between both parameters regardless of the $M_\star$ value 
assumed. Lower $\alpha_{\rm CO}$ values imply higher dark matter fractions. We 
note that if we assume the \citet{Accurso2017}'s $\alpha_{\rm CO}$ value, we obtain 
$f_{\rm DM} \sim 0.3 \pm 0.13$. Although SHiZELS-19 has a metallicity consistent with being solar, 
its ISM morphology and kinematics departs strongly from the ISM conditions observed 
in local galaxies. The high molecular gas velocity dispersion values (Table~\ref{tab:table3})
observed for this system suggest that SHiZELS-19 should have a dense ISM \citep{Papadopoulos2012} 
which may lower its CO-to-H$_2$ conversion factor value \citep{Bolatto2013}. 
As the \citet{Accurso2017}'s parametrization does not consider the ISM density effects, 
its $\alpha_{\rm CO}$ value should be considered as an upper limit. This is also consistent 
with the $\alpha_{\rm CO}$ upper limit derived from the dynamical mass estimate within 
the CO half-light radius (see Appendix~\ref{appendix:B}, for more details.). Thus, in the remaining 
of this work, we use the \citet{Narayanan2012}'s parametrization as it does consider the 
ISM density effect in the estimation of $\alpha_{\rm CO}$.

We find $f_{\rm DM} \approx 0.59\pm0.10$ for SHiZELS-19. This 
value is consistent with the dark matter fraction predicted for disk-like galaxies at similar 
redshift range and stellar mass content from hydrodynamical simulations \citep{Lovell2018}. 
From our Bayesian approach, we find 3-$\sigma$ range boundaries of $\sim 0.31-0.70$ for 
the $f_{\rm DM}$ value. On the other hand, if we consider the extreme case of a stellar mass 
underestimated by a factor of two \citep{SS2018}, then the 3-$\sigma$ range boundaries 
correspond to $\sim 0.20-0.64$.

To determine if we need to include the H{\sc i} content in our analysis, we note that
in local spirals the transition between a H$_2$- to H{\sc i}-dominated ISM 
($\Sigma_{\rm H_2}\approx \Sigma_{\rm HI}$) occurs at a gas surface density of
$\Sigma_{\rm gas} \sim 12 \pm 6$\,$M_\odot$\,pc$^{-2}$ \citep{Leroy2008}. In 
contrast, from our spatially resolved CO(2-1) observations, we derive an average 
$\Sigma_{\rm H_2} \sim 220\pm166$\,$M_\odot$\,pc$^{-2}$ value from a tilted ring
centred at the the same radius at which $V_{\rm rot,tap}$ was calculated. 
This suggests that the H{\sc i} mass content within a radius of $\approx$6\,kpc 
is likely to be negligible compared to $M_{\rm H_2}$ and therefore, our estimated 
$f_{\rm DM}$ value may be a good approximation. Thus, we suggest that SHiZELS-19 is 
a `typical' star-forming galaxy which may have a considerable dark matter content.

The dark matter fraction obtained for SHiZELS-19 is consistent with the values reported by 
\citet{Tiley2018}, but in tension with the conclusions reported from \citet{Genzel2017} 
and \citet{Lang2017}. These three studies rely primarily on the analysis of the stacked 
rotation curve constructed from normalized individual velocity curves from galaxies in the 
$0.6\lesssim z \lesssim 2.6$ redshift range. However, the discrepancy between the 
obtained $f_{\rm DM}$ values from these studies seems to be driven by the way in which 
the velocity curves are normalized. In \citet{Genzel2017} and \citet{Lang2017} works they 
normalize the individual velocity curves in both, radial extension through the turn-over 
radius and velocity amplitude through the velocity at the turn-over radius (see \citealt{Lang2017} 
for more details). This normalization procedure tends to favour the contribution of the systems 
with low $V_{\rm rot}/\sigma_v$ values to the stacked rotation curve at longer radii. This bias 
seems to be produced by the smaller turn over radius values presented in those galaxies 
which acts as a `zoom in' scaling factor \citep{Tiley2018}. On the other hand, \citet{Tiley2018} 
normalize the individual velocity curves by the stellar light disk-scale radius and also velocity. 
In this case, galaxies with different $V_{\rm rot}/\sigma_v$ values contribute more uniformly 
to the shape of the stacked rotation curve. Taking this into account, we note that SHiZELS-19 is a 
galaxy with $V_{\rm rot}/\sigma_v \sim 1.0-1.3$, favouring the scenario in which the conclusions 
presented in \citet{Genzel2017} and \citet{Lang2017} studies may be biased.

\begin{table}
	\centering
	\setlength\tabcolsep{4pt}
	TABLE 4: SHiZELS-19 FINAL PARAMETERS\\
    	\caption{\label{tab:table4}                                                   
	Summary of the SHiZELS-19 galaxy parameters derived in this work using the kinematic modelling. 
	The $f_{\rm DM}$, $\Sigma_{\rm H_2}$ and $\tau_{\rm dep}$ values are computed by considering 
	the \citet{Narayanan2012}'s CO-to-H$_2$ conversion factor (see \S~\ref{sec:dyn_mass}).}
	\begin{tabular}{lc} 
		\hline
		\hline
	    $f_{\rm DM}$ & $0.59 \pm 0.10$ \\
        $\log_{10} \Sigma_{\rm SFR}$ ($M_\odot$\,kpc$^{-2}$) & $-0.5 \pm0.3$ \\
        $\log_{10} \Sigma_{\rm H_2}$ ($M_\odot$\,pc$^{-2}$) & $3.0\pm0.6$ \\
        $\tau_{\rm dep}$ (Gyr) & 2.3 $\pm$ 1.2 \\
		\hline                                                                                        
	\end{tabular}
\end{table}

\subsection{The Kennicutt-Schmidt law at $\sim$kpc-scales}
\label{sec:sf_law}

First proposed by \citet{Schmidt1959} and extended by \citet{Kennicutt1998a, 
Kennicutt1998b}, the Kennicutt-Schmidt law is an observational power-law 
relationship between the galaxy star formation rate surface density ($\Sigma_{\rm SFR}$) 
and the gas surface density. It describes how efficiently galaxies turn their gas 
content into stars. For local galaxies, this correlation is well-fitted by an exponent 
of $N=1.4$ \citep{Kennicutt1998b}. 

Since then, latter studies have found that $\Sigma_{\rm SFR}$ is better correlated with
the molecular gas surface density ($\Sigma_{\rm H_2}$) rather than $\Sigma_{\rm gas}$ 
(e.g. \citealt{Bigiel2008, Leroy2008}). At first order, local disk-like galaxies show a linear 
correlation between both surface density quantities ($\Sigma_{\rm SFR} \propto \Sigma_{\rm H_2}$), 
with a median depletion time ($\tau_{\rm dep} \equiv \Sigma_{\rm H_2}/\Sigma_{\rm SFR}$)
of $\sim$2.2\,Gyr (e.g. \citealt{Leroy2013}). Second order departures from this relationship have
also been found (e.g. \citealt{Saintonge2012,Utreras2016}), although these effects may be related to 
systematic errors behind the estimation of the molecular gas content and/or local nuclear 
starburst activity \citep{Leroy2013}.

\begin{figure*}
\includegraphics[width=\columnwidth]{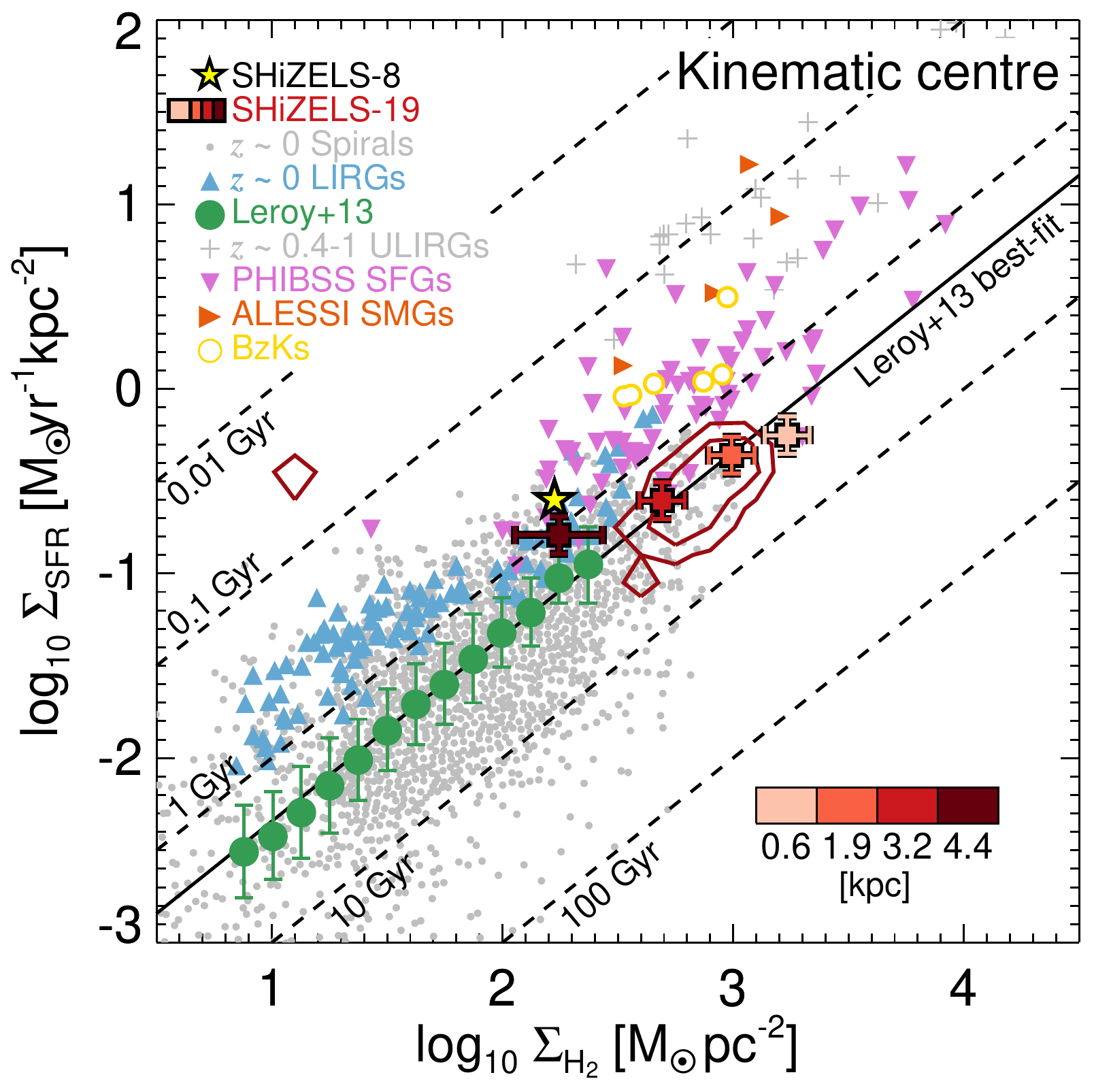}
\includegraphics[width=\columnwidth]{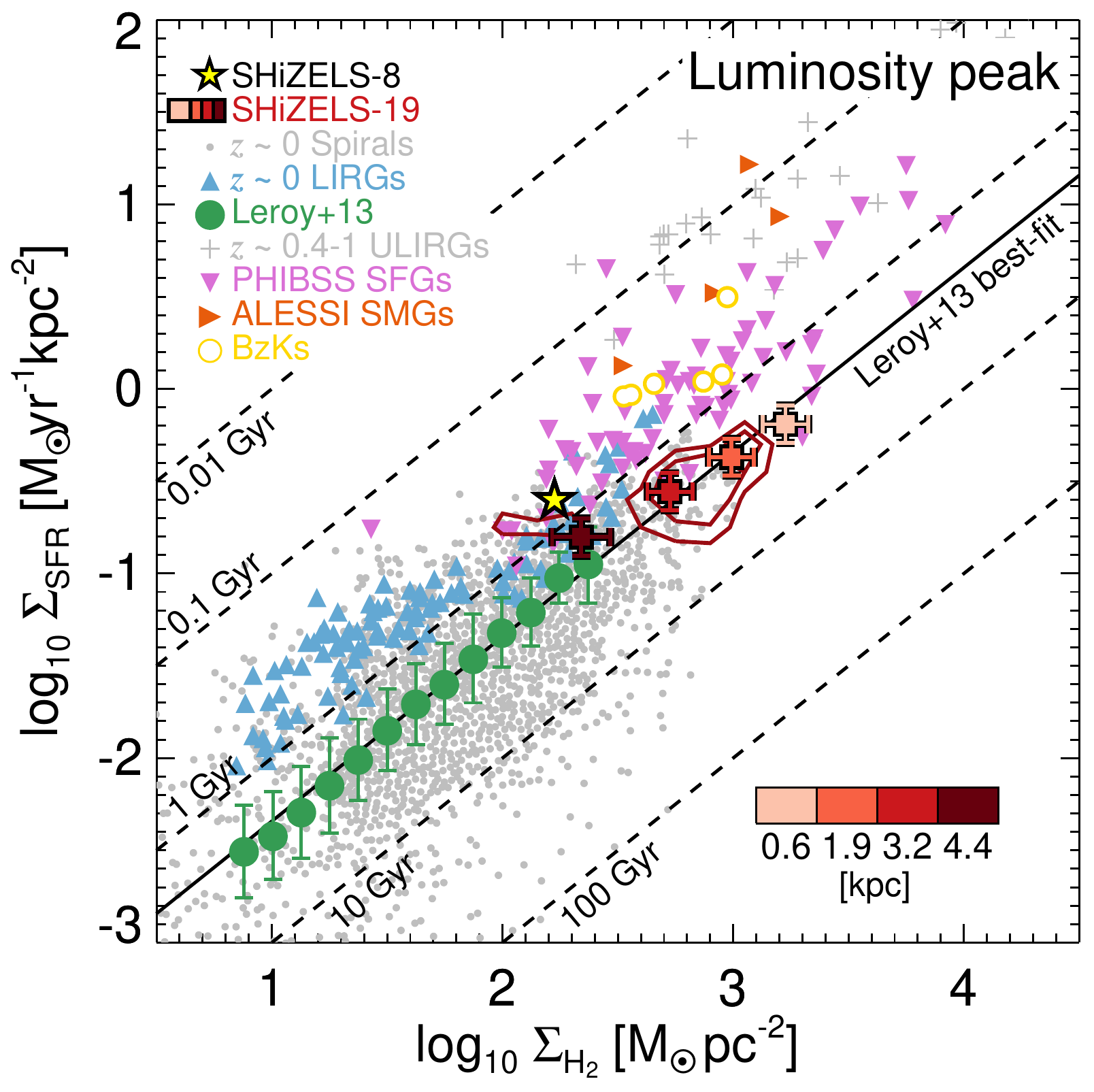}
\caption{ \label{fig:ks_plot}
\textit{Left:} $\Sigma_{\rm SFR}$ against $\Sigma_{\rm H_2}$ for SHiZELS-8 and SHiZELS-19 galaxies 
compared with spatially-resolved local galaxy observations in the literature. For SHiZELS-8 we just show
a global estimate given the limitations of our marginally-detected CO observation. For SHiZELS-19 we centre the CO(2-1) 
and H$\alpha$ two-dimensional intensity maps by using the best-fit kinematic centre. The 
red contours show the 50th and 90th percentile levels of the pixel-by-pixel distribution. The 
colour-coded squares represent the $\Sigma_{\rm SFR}- \Sigma_{\rm H_2}$ values calculated 
within tilted rings of 0$\farcs$15 ($\sim 1.2$\,kpc) thickness at the radius indicated by the 
colour bar. The `$z\sim0$ Spirals' sample consist in observations of local disks taken 
from \citet{Kennicutt2007, Blanc2009, Rahman2011, Rahman2012} at spatial resolutions 
between $\sim 0.2-1$\,kpc. The `$z\sim0$ LIRGs' values consist in $\sim 1$\,kpc scale 
observations of two galaxies \citep{Espada2018}. The green circles show the median trend 
observed in the HERACLES survey \citep{Leroy2013} at $\sim$kpc-scales and the black line 
represents the best-fit for those median values. The error bars represent the 1-$\sigma$ 
uncertainty. We also present galaxy-integrated estimates of ULIRGs ($z \sim 0.4-1$; \citealt{Combes2013a}), 
four SMGs taken from the ALESS survey ($z\sim 2.0-2.9$; \citealt{Calistro2018}), `typical' 
star-forming galaxies observed at similar redshifts \citep{Tacconi2013} and five $BzK$ galaxies 
($z \sim 1.5$; \citealt{Daddi2010}). The dashed lines indicate fixed $\tau_{\rm dep}$ values. 
\textit{Right:} The same plot as the showed in left panel, but now the CO(2-1) and H$\alpha$ 
two-dimensional intensity maps are centred by using the CO and H$\alpha$ luminosity peaks.
}
\end{figure*}

In Fig.~\ref{fig:ks_plot} we present the star formation activity of SHiZELS-8 and SHiZELS-19 in the context of the 
$\Sigma_{\rm SFR} - \Sigma_{\rm H_2}$ relation. We compare with several local galaxy samples observed 
at similar spatial scales and galactic averages of galaxies observed at similar redshifts. 
Briefly, the `$z \sim 0$ spirals' sample is composed by high spatial resolution ($\sim 0.2-1$\,kpc) 
observations of small galactic regions taken from \citet{Kennicutt2007, Blanc2009, Rahman2011, 
Rahman2012}. The `$z \sim 0$ LIRGs' sample consist in observations of the NGC3110 and 
NGC232 galaxies observed at $\sim 1$\,kpc scale \citep{Espada2018}. Both starburst systems 
have SFR$\sim 20$\,$M_\odot$\,yr$^{-1}$, i.e. comparable with the value reported for SHiZELS-19
(SFR=23\,$M_\odot$\,yr$^{-1}$). NGC3110 is a barred Sb galaxy interacting with a minor 
companion (mass ratio $\sim$14:1), whilst NGC232 corresponds to a barred Sa galaxy
which presents a bright compact nuclear region (see \citealt{Espada2018}, for more details). 
We also compare with the median trend observed for a sub-sample of 30 local galaxies 
taken from the HERA CO Line Extragalactic Survey (HERACLES; \citealt{Leroy2013}). These 
data also consist in $\sim 1$\,kpc scale observations of the galactic ISM. The high redshift observations 
consist in galaxy-integrated estimates from ULIRGs ($z \sim 0.4-1$; \citealt{Combes2013a}),
four SMGs taken from the ALESS survey ($z\sim 2.0-2.9$; \citealt{Calistro2018}), 
`typical' star-forming galaxies observed at $z \sim 1-2.5$ taken from the PHIBSS survey 
\citep{Tacconi2013} and five $BzK$ galaxies ($z \sim 1.5$) presented in \citet{Daddi2010}. For 
the ALESS SMGs we calculate the surface density quantities following \citet{Tacconi2013}.

Given our marginally-detected CO observation for SHiZELS-8, we just 
plot a galactic average estimation ($\log_{10} \Sigma_{\rm SFR} \approx -0.61 \pm 0.07$\,$M\odot$\,yr$^{-1}$\,kpc$^{-2}$; 
$\Sigma_{\rm H_2} \sim 2.23 \pm 0.08$\,$M\odot$\,pc$^{-2}$). However, we caution that in this particular case, $\Sigma_{\rm H_2}$ is a rough 
estimation as we can not constrain the SHiZELS-8's CO spatial distribution accurately. For SHiZELS-19, we use the Interactive Data Language (IDL)
procedure \textsc{hastrom} to align the two-dimensional fields using as a reference point the 
kinematic centre (left panel) and the luminosity peak position (right panel) from H$\alpha$ and 
CO $\sim$kpc-scale observations. For this galaxy, we derive a median $\log_{10} \Sigma_{\rm SFR} =-0.5 \pm0.3$\,$M_\odot$\,yr$^{-1}$\,kpc$^{-2}$ 
and $\log_{10} \Sigma_{\rm H_2} = 3.0\pm0.6$\,$M_\odot$\,pc$^{-2}$ values. These estimations indicate that 
SHiZELS-19 has a somewhat denser ISM compared with local star-forming galaxies. On the other hand, the 
median $\Sigma_{\rm H_2}$ value is consistent with molecular surface density estimations from galaxy-integrated 
observations of $BzK$ and `typical' star-forming galaxies at similar redshifts \citep{Daddi2010,Tacconi2013},
but SHiZELS-19 presents lower $\Sigma_{\rm SFR}$ values compared to these systems.

We derive a median $\tau_{\rm dep}=2.3 \pm 1.2$\,Gyr for this galaxy, with the pixel-by-pixel 
distribution between $\sim 0.003-5$\,Gyr. SHiZELS-19 presents a median depletion time consistent 
with the best-fit $\tau_{\rm dep}=2.2\pm0.28$\,Gyr reported in \citet{Leroy2013} for the median 
trend observed in local galaxies at similar spatial resolution.

In the left panel of Fig.~\ref{fig:ks_plot} we show the $\tau_{\rm dep}$ values calculated from
tilted rings constructed from the two-dimensional best-fit model and centred at the kinematic 
centre. At first order, we find the same trend suggested from the average $\tau_{\rm dep}$ 
values. But, at second order, we note that the depletion times vary from $\sim 1.0 \pm 0.3$\,Gyr
in the outer ring ($\approx$4.4\,kpc) to $\sim 2.9 \pm 0.2$\,Gyr in the central kpc of this galaxy, suggesting 
an apparent decrease in the star formation efficiency (SFE$\equiv \tau_{\rm dep}^{-1}$) towards 
the galactic centre in SHiZELS-19. This is in contradiction with second order effects found in galaxies 
in the local Universe. Possible variations of the CO-to-H$_2$ conversion factor through a radial 
dependence of the dust-to-gas ratio optical depth or gas excitation or nuclear starburst activity in 
galactic centres favour the opposite $\tau_{\rm dep}$ correlation with galactic radius 
(e.g. \citealt{Sandstrom2013, Leroy2013}). However, by using the [N{\sc ii}/H$\alpha$] ratio as a 
proxy of the metallicity gradient \citep{Pettini2004}, we find a $\alpha_{\rm CO}$ radial profile 
consistent with being flat (see Appendix~\ref{appendix:A}).
 
On the other hand, although the CO and H$\alpha$ maps are smooth, the H$\alpha$ best-fit 
kinematic centre does not coincide exactly with the H$\alpha$ luminosity peak as it does in the CO 
observations. Indeed, the projected distance between both centres is $\sim 0\farcs11$, 
i.e, slightly lower than the spatial resolution of the observations ($\approx 0\farcs$15). Thus, possible 
inaccuracies of our best-fit kinematic centres given by the limited spatial resolution of our observations 
may lead to the apparent outward decrease of the $\tau_{\rm dep}$ values obtained from the tilted rings.
In order to explore this possibility, in the right panel of Fig.~\ref{fig:ks_plot} we show the $\tau_{\rm dep}$ 
values calculated from tilted rings constructed from the two-dimensional best-fit model but centred at
the luminosity peak. For this case, the $\tau_{\rm dep}$ values vary from $\sim 1.3 \pm 0.3$\,Gyr in the 
outer ring to $\sim 2.5\pm0.1$ in the inner ring. The increase of the $\tau_{\rm dep}$ values towards the
galactic centre still remains. 

The suppression of the star formation in the molecular gas by dynamical effects is a possibility. For 
example, a morphological quenching scenario in which the bulge stabilises the molecular gas,
preventing the star-formation activity but not destroying the gas may explain the observed $\tau_{\rm dep}$ 
trend with galactocentric radius (e.g. \citealt{Martig2009,Saintonge2011}). However, this scenario is 
unlikely as the S\'ersic index measured for SHiZELS-19 ($n \sim 1$; \citealt{Gillman2019}) indicates that 
this galaxy is consistent with being a disk-like galaxy with no prominent bulge component. Galaxies with 
a prominent bulge component tend to show S\'ersic index values deviated from unity \citep{Lang2014}. 
On the other hand, \citet{Schreiber2016} found that the increase of $\tau_{\rm dep}$ towards the central 
galactic zone in massive systems ($M_\star \sim 10^{11}$\,$M\odot$) seems to be independent of the 
possible mass growth of the bulge component as disk-dominated galaxies tend to present the same 
$\tau_{\rm dep}$ trend with radius.

Another possible effect that adds uncertainty to the calculated $\tau_{\rm dep}$ values is a potential spatial 
variation of the H$\alpha$ extinction. We have used an $A_{\rm H\alpha}$ correction constant across the 
galactic disk, but an underestimated galactic extinction in the galactic centre may lower the observed 
$\tau_{\rm dep}$ values in the central kpc zone therefore, producing the observed trend.
An increase of the H$\alpha$ extinction towards the galactic centre is consistent with findings of 
$A_{\rm H\alpha}$ being correlated with stellar mass surface density \citep{Hemmati2015} or 
the presence of compact density starbursts (e.g. \citealt{Simpson2015, Hodge2016,Hodge2019})

In order to explore the effects of the global galaxy kinematics in the global star formation activity,
we compute the orbital timescale ($\tau_{\rm orb}= 2 \pi R/ V_{\rm rot}$) to be compared 
it with the median depletion timescale (e.g. \citealt{Kennicutt1998b, Daddi2010}). By following the 
analysis of \citet{Daddi2010}, we choose $R$ to be equal to three times the half-light radius. 
Although this assumes that the rotation curve remains flat beyond two half-light radius (the 
radius at which $V_{\rm rot}$ was estimated), this seems to be a reasonable assumption (see 
Fig.~\ref{fig:COS30_tapered} \& \citealt{Tiley2018}). Thus, we obtain $\tau_{\rm orb} = 256\pm22$\,Myr 
and $\tau_{\rm dep}/\tau_{\rm orb} \sim 9 \pm 5$. We find that SHiZELS-19 converts $\sim$10\% 
of its available gas into stars per orbit. This is consistent with the average value found for local 
galaxies by \citet{Kennicutt1998b} and with galaxy-integrated studies of $BzK$ galaxies at similar 
redshifts \citep{Daddi2010}. Therefore, on average, SHiZELS-19 is a galaxy which follows a similar 
star formation law to that seen in local spiral galaxies, although in denser environments.

We should stress, however, that our conclusions are highly dependent on the assumed 
$\alpha_{\rm CO}$ value (Table~\ref{tab:table4}) and its variation with radius. We have used the CO-to-H$_2$ 
conversion suggested by \citet{Narayanan2012} in order to consider possible variations in 
the average ISM metallicity and density (see also Appendix~\ref{appendix:A}). However, spatially 
resolved observations of the dust content are desirable as these may help to constrain the 
$\alpha_{\rm CO}$ value through a dust-to-gas-to ratio based approach \citep{Leroy2013, Sandstrom2013}.

Our work opens the possibility to perform morpho-kinematic analysis of high-redshift
galaxies at $\sim$kpc-scales using two different ISM tracers, but we stress that 
more observations of `typical' galaxies are needed to understand the impact of 
local or global galactic properties on the star formation activity in high-redshift systems.

\section{Conclusions}

We present new ALMA Cycle-5 observations tracing the CO(2-1) emission line from two `typical' star-forming 
galaxies at $z\sim1.47$. These observations were designed to deliver spatially resolved observations of the molecular 
gas content on $\sim$kpc-scales. We combine our ALMA observations with the previous H$\alpha$ SINFONI AO-aided 
observations (\citealt{Swinbank2012a,Molina2017, Gillman2019}) in order to study and compare the ionized and 
molecular gas dynamics.

One of our targets, SHiZELS-8, is marginally detected only in the 2000k$\lambda$ tapered datacube ($0\farcs5 \sim 4.3$\,kpc 
spatial resolution). For this system the H$\alpha$ and CO dynamics show that both ISM components 
rotate in the same direction but have position angles offset by $100-120$\,deg. This suggests that SHiZELS-8 is a 
dynamically perturbed system consistent with its previously observed flat metallicity gradient \citep{Swinbank2012a}.
This finding suggests that `main-sequence' galaxies at high-redshift are not exclusively part of a well-behaved
morpho-kinematic disk-like population (e.g. \citealt{Elbaz2018}).

For the other target, SHiZELS-19, we find a good agreement between the CO and H$\alpha$ spatial extent 
($r_{\rm 1/2,H\alpha}/ r_{\rm 1/2,CO} \sim 1.07 \pm 0.09$) and dynamics at $\sim$kpc-scales (Fig.~\ref{fig:maps}). 
From both ISM phases we derive $V_{\rm rot}/\sigma_v \sim1$ (Table~\ref{tab:table3}). By performing a kinemetry 
analysis we classify SHiZELS-19 as a `non-regular rotator' \citep{vandenSande2017}. The kinematic analysis suggests 
that the CO and H$\alpha$ observations are tracing the same galactic kinematics in agreement with previous studies 
of massive galaxies at similar redshift range (e.g. \citealt{Ubler2018,Calistro2018}).

We estimate the total mass budget of the SHiZELS-19 galaxy by assuming a galactic thick-disk geometry \citep{Burkert2010}
and \citet{Narayanan2012}'s CO-to-H$_2$ conversion factor. From the SHiZELS-19 2000k$\lambda$ datacube we are 
able to trace the CO emission up to $\approx 6$\,kpc (Fig.~\ref{fig:COS30_tapered}), finding a dark matter fraction of 
$f_{\rm DM} = 0.59\pm0.10$ within this aperture. By applying a MCMC technique to sample the posterior PDF 
and take into account the parameter uncertainties (Fig.~\ref{fig:distributions}; e.g. \citealt{Calistro2018}) we estimate a 
$f_{\rm DM}$ 3-$\sigma$ error range of $\sim$0.31$-$0.70. The dark matter fraction value is in agreement with 
hydrodynamical simulations of disk-like galaxies with similar stellar mass \citep{Lovell2018} and the average dark matter 
fraction suggested by the stacked rotation curve analysis of galaxies at similar redshift range \citep{Tiley2018}. Thus, we 
conclude that SHiZELS-19 is a `typical' star-forming galaxy at $z\sim1.47$ harbour in a non-negligible amount of dark matter.

By using two-dimensional modelling, we study the star formation activity observed in the SHiZELS-19
galaxy at $\sim$kpc-scales. We derive a median $\tau_{\rm dep}=2.3 \pm 1.2$\,Gyr. This median 
value is consistent with the typical value observed in local galaxies at similar spatial scales 
($\tau_{\rm dep}=2.2 \pm 0.28$\,Gyr; \citealt{Leroy2013}) and consistent with the the large scatter 
presented in the $z\sim0$ spirals galaxy observations (Fig.~\ref{fig:ks_plot}), suggesting that `typical' 
high-redshift galaxies (at $z \sim 1.47$) with denser ISM still follow the same star-formation law.

\section*{Acknowledgements}

We thank to the anonymous referee for her/his careful read of the manuscript and helpful comments and suggestions.
J. M.~acknowledges the support given by CONICYT Chile (CONICYT-PCHA/Doctorado-Nacional/2014-21140483).
E. I.~acknowledges partial support from FONDECYT through grant N$^\circ$\,1171710.
A.E.~acknowledges support from CONICYT project Basal AFB-170002 and Proyecto Regular FONDECYT (grant 1181663).
I.R.S. and A.M.S.~acknowledge the support from STFC (ST/P000541/1).
This paper makes use of the following ALMA data: ADS/JAO.ALMA\#2018.1.01740.S, ADS/JAO.ALMA\#2012.1.00402.S
and ADS/JAO.ALMA\#2015.1.00862.S. ALMA is a partnership of ESO (representing its member states), NSF (USA) and 
NINS (Japan), together with NRC (Canada), NSC and ASIAA (Taiwan), and KASI (Republic of Korea), in co- operation 
with the Republic of Chile. The Joint ALMA Observatory is operated by ESO, AUI/NRAO and NAOJ.





\bibliographystyle{mnras}



\appendix

\section{$\alpha_{\rm CO}$ radial profile}
\label{appendix:A}

\begin{figure}
\includegraphics[width=\columnwidth]{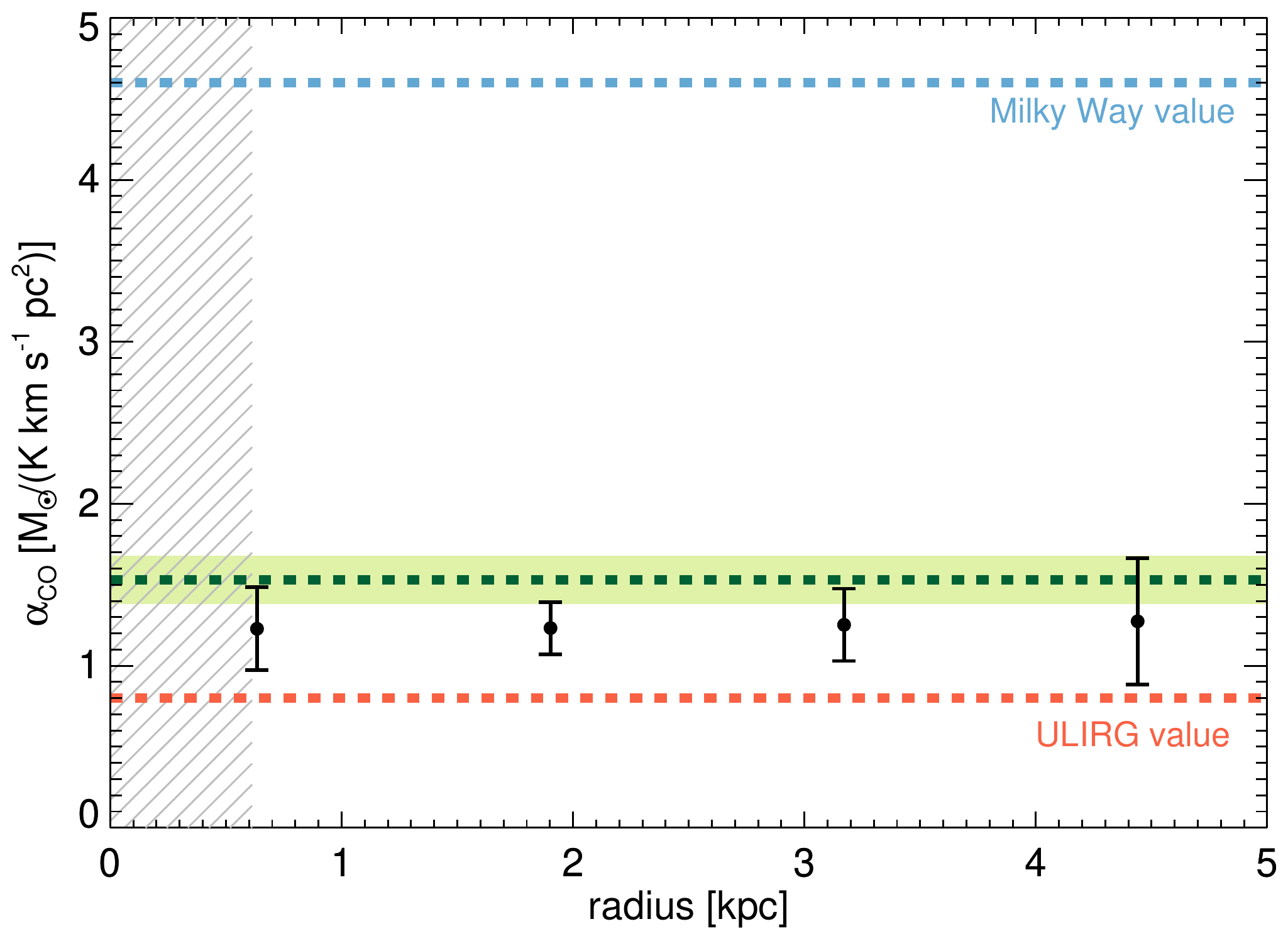}
\caption{ \label{fig:aco_prof}
CO-to-H$_2$ conversion factor gradient across SHiZELS-19 from spatially-resolved measurements and 
as a function of the galactocentric radius derived from the best-fit kinematic model. The green-dashed line represent 
the galactic $\alpha_{\rm CO}$ value derived from the tapered (2000\,k$\lambda$) map and the green-shaded region 
show the 1-$\sigma$ uncertainty. The blue- and red-dashed lines indicate the $\alpha_{\rm CO}$ values usually
adopted for the Milky Way and ULIRG like systems. The grey-dashed area represents the resolution element 
radial extent. We find a flat $\alpha_{\rm CO}$ profile.
}
\end{figure}

Throughout this work we have used a simple CO-to-H$_2$ conversion factor to estimate the molecular 
gas content in SHiZELS-8 and SHiZELS-19 galaxies (\S~\ref{sec:molgas_cont}). Thus, we have assumed that 
there is no significant radial variation of the $\alpha_{\rm CO}$ value across each galactic disk. In order to 
test this assumption, we calculate the CO-to-H$_2$ conversion factor radial profile. This can only be done 
for the SHiZELS-19 galaxy since we were not able to obtain spatially resolved CO observations for SHiZELS-8. 
In Fig.~\ref{fig:aco_prof} we show the $\alpha_{\rm CO}$ as a function of the galactocentric radius. 
It was calculated by using the \citet{Narayanan2012}'s parametrization with the CO surface brightness radial 
profile and metallicity gradient as input values. We find an $\alpha_{\rm CO}$ gradient consistent with being 
flat. This is mainly produced by the sub-linear dependence of the CO-to-H$_2$ conversion factor 
with respect to $\Sigma_{\rm CO}$ and metallicity in the \citet{Narayanan2012}'s parametrization. Although 
SHiZELS-19 has a negative metallicity gradient \citep{Molina2017} it does not vary enough in order to increase 
the $\alpha_{\rm CO}$ value at larger radii.

We note that the $\alpha_{\rm CO}$ radial profile values are slightly lower but still consistent within
1-$\sigma$ uncertainties with the galactic average CO-to-H$_2$ conversion factor value calculated 
from the tapered (2000\,k$\lambda$) map. This is expected as the low spatial resolution data-cube
is able to trace CO(2-1) emission from the outskirts of the galaxy where the CO surface brightness is 
lower and the molecular gas has low metallicity compared to the inner parts. Both effects favour the 
increase of the average $\alpha_{\rm CO}$ value.

It is worth to mention that by considering the large variety of metallicity gradients observed in high
redshift galaxies (e.g. \citealt{Queyrel2012, Swinbank2012a, Molina2017}), this result may be 
particularly applicable to SHiZELS-19 and it might not be used as typical property for 
the bulk population.

\section{$\alpha_{\rm CO}$ upper limit from dynamics}
\label{appendix:B}

In \S~\ref{sec:dyn_mass} we have assumed the \citet{Narayanan2012}'s parametrization to estimate the CO-to-H$_2$
conversion factor. Using this $\alpha_{\rm CO}$ value coupled with our dynamical mass calculus, we have constrained 
the dark matter content in the SHiZELS-19 galaxy. We have used the \citet{Narayanan2012}'s parametrization in detriment of
\citet{Accurso2017}'s parametrization as the second is likely to be an upper limit for the CO-to-H$_2$ conversion factor as
it does not consider the gas surface density effects \citep{Bolatto2013}. In order to confirm this assumption we use the 
dynamical mass calculus to constrain the CO-to-H$_2$ conversion factor (e.g. \citealt{Tacconi2008}). 

We repeat the analysis done in \S~\ref{sec:dyn_mass}, but now we calculate the total and stellar mass content within one 
CO half-light radius. The stellar mass within this radius is estimated by assuming an exponential stellar surface density 
profile, as suggested by the best-fitted S\'ersic profile presented in \citet{Gillman2019} for the \textit{HST}-F160W broad-band image. 
We caution, however, that this calculus also assumes a constant mass-to-light ratio across the SHiZELS-19 galactic disk. 
We calculate the thick-disk dynamical mass within one $r_{\rm 1/2,CO}$ by using the $\sim$kpc-scale kinematic CO 
observations (Fig.~\ref{fig:maps}).

Initially we just constrain the $\alpha_{\rm CO}$ lower limit value by imposing that the CO emission should be optically 
thick ($\alpha_{\rm CO} \gtrsim 0.34$; \citealt{Bolatto2013}). We do not assume any dark matter content as we allow 
that the MCMC technique samples $\alpha_{\rm CO}-f_{\rm DM}$ phase-space and fully considers the parameter 
degeneration introduced in Eq.~\ref{eq:f_dm}.

\begin{figure}
\includegraphics[width=\columnwidth]{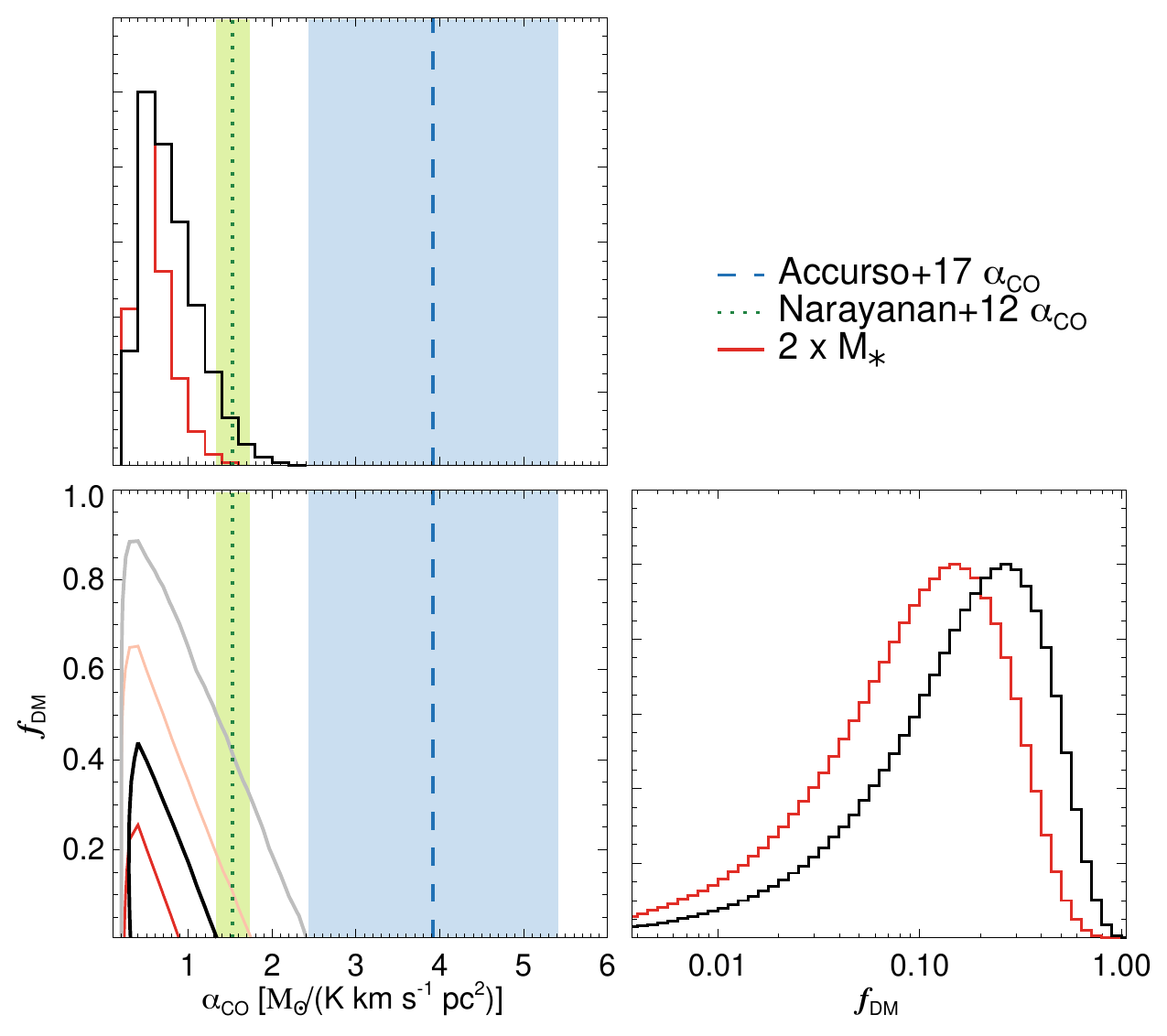}
\caption{ \label{fig:appendix_B}
One- and two- dimensional posterior PDFs of the $f_{\rm DM}$ and $\alpha_{\rm CO}$ parameters 
estimated by considering the total mass content within one CO half-light radius for SHiZELS-19. The data 
is colour coded in the same way as Fig.~\ref{fig:distributions}. This suggests an 
$\alpha_{\rm CO}$ upper limit of  2.4\,$M_\odot$\,(K\,km\,s$^{-1}$\,pc$^2$)$^{-1}$ in the case of 
negligible dark matter content within this radius. This result rules out the \citet{Accurso2017}'s CO-to-H$_2$ 
conversion factor suggested for SHiZELS-19 by the 3-$\sigma$.
}
\end{figure}

In Fig.\ref{fig:appendix_B} we show the one- and two- dimensional posterior PDFs of the $\alpha_{\rm CO}$ and $f_{\rm DM}$ 
parameters. As in \S~\ref{sec:dyn_mass}, we find that higher $\alpha_{\rm CO}$ values imply lower dark matter fractions.
In the case of negligible central dark matter content within $r_{\rm 1/2,CO}$, we find an $\alpha_{\rm CO}$ 
upper limit of 1.3(2.4)\,$M_\odot$\,(K\,km\,s$^{-1}$\,pc$^2$)$^{-1}$ by considering 1-(3-)\,$\sigma$ uncertainties.

This analysis suggests that the \citet{Accurso2017}'s parametrization overestimates the CO-to-H$_2$ conversion factor in 
SHiZELS-19 as this value is beyond the 3-$\sigma$ range derived from the $\alpha_{\rm CO}$ PDF. Meanwhile, the CO-to-H$_2$
conversion factor estimated by assuming the \citet{Narayanan2012}'s parametrization is consistent within 1-$\sigma$ uncertainties.
We note that an assumed $\alpha_{\rm CO} \gtrsim 1$ implies that SHiZELS-19 may be baryon dominated ($f_{\rm DM}<0.5$) in its 
central zone, albeit dark matter dominated in its outskirts (\S~\ref{sec:dyn_mass}; see also \citealt{Tiley2018}). This is consistent with 
the `compaction' scenario (e.g. \citealt{Dekel2014,Zolotov2015}) in which the baryonic matter can cool and condense more efficiently 
than the collisionless dark matter, and thus, falling into the centre of the dark matter halo where they concentrate.

\bsp	
\label{lastpage}
\end{document}